%% file: 00-main.tex
\documentclass[sigconf,sigconf]{acmart}

\setcitestyle{square}

\input{01-pre}

\input{02-siggraph}



\title{High-Order Incremental Potential Contact for Elastodynamic Simulation on Curved Meshes}

\author{Zachary Ferguson}
\email{zfergus@nyu.edu}
\affiliation{%
  \institution{New York University}
  \country{USA}
}

\author{Pranav Jain}
\email{pranav.jain@nyu.edu}
\affiliation{%
  \institution{New York University}
  \country{USA}
}

\author{Denis Zorin}
\email{dzorin@cs.nyu.edu}
\affiliation{%
  \institution{New York University}
  \country{USA}
}

\author{Teseo Schneider}
\email{teseo@uvic.ca}
\affiliation{%
  \institution{University of Victoria}
  \country{Canada}
}

\author{Daniele Panozzo}
\email{panozzo@nyu.edu}
\affiliation{%
  \institution{New York University}
  \country{USA}
}

\AtBeginMaketitle{%
\let\oldAtBeginDocument\AtBeginDocument%
\renewcommand\AtBeginDocument[1]{#1}
  \input{02-siggraph}
\let\AtBeginDocument\oldAtBeginDocument%
}

\begin{document}

\input{03-abstract}
\input{04-teaser}

\maketitle

\renewcommand{\shorttitle}{High-Order IPC for Elastodynamic Simulation on Curved Meshes}

\acresetall %

\input{10-introduction}

\input{20-related}

\input{30-background}
\input{40-method}
\input{50-results}

\input{60-concluding}

\input{65-acknowledgments}

\bibliographystyle{ACM-Reference-Format}
\bibliography{99-bib}
\newpage
\clearpage
\input{70-figures}

\newpage
\clearpage
\end{document}

%% file: 01-pre.tex
\usepackage{acmart-taps}

\usepackage[utf8]{inputenc}
\usepackage{algorithm}
\usepackage{algorithmic}

\usepackage{natbib}
\usepackage{subcaption}

\usepackage{xcolor}
\usepackage{ifthen}
\usepackage{siunitx}
\usepackage[capitalize,noabbrev]{cleveref}

\usepackage{multirow}
\usepackage{makecell} %

\newcommand{\real}{\mathbb{R}}

\newcommand{\vecspace}{\Theta}
\newcommand{\W}{W}
\aptLtoXcmd{\newcommand{\M}{\mathcal{M}}}{%
\newcommand{\M}{{
  \mathchoice{\mathcal{M}}{\mathcal{M}}{\scriptscriptstyle\mathcal{M}}{\mathcal{M}}
}}}
\newcommand{\linearM}{\widetilde \M}
\aptLtoXcmd{\renewcommand{\S}{\mathcal{S}}}{%
\renewcommand{\S}{{
  \mathchoice{\mathcal{S}}{\mathcal{S}}{\scriptscriptstyle\mathcal{S}}{\mathcal{S}}
}}}

\newcommand{\sspan}[1]{\text{span}\{#1\}}
\newcommand{\dhms}[4]{#1d~#2h~#3m~#4s}
\newcommand{\hms}[3]{#1h~#2m~#3s}
\newcommand{\ms}[2]{#1m~#2s}
\aptLtoXcmd{\newcommand{\basis}{\varphi}}{\newcommand{\basis}{\text{\larger[1]$\varphi$}}}
\DeclareMathOperator*{\argmin}{\arg\!\min}

\let\vec\mathbf
\newcommand{\transposeSymbol}{\mathstrut\mathbf{\top}}
\newcommand{\transpose}[1]{#1^{\transposeSymbol}}

\newcommand{\aabb}{\boxdot}

\usepackage{acronym}
\newacro{IP}{incremental potential}
\newacro{IPC}{Incremental Potential Contact}
\newacro{CIPC}[C-IPC]{Codimensional \ac{IPC}}
\newacro{NURBS}{non-uniform rational B-spline}
\newacro{IGA}{isogeometric analysis}
\newacro{DOF}{degrees of freedom}
\newacro{FE}{finite element}
\newacro{FEM}{finite element method}
\newacro{CCD}{continuous collision detection}
\newacro{LCP}{linear complementarity problem}
\newacro{LBFGS}[L-BFGS]{Limited-memory BFGS}
\newacro{HPC}{high performance computing}
\newacro{HO}{high-order}

\newcommand{\copyrightmodel}[2]{\begin{flushright}{\scriptsize Model \copyright #1 under #2.}\end{flushright}}

\newcommand{\figname}[1]{\emph{#1}}

%% file: 02-siggraph.tex
\acmSubmissionID{231}

\setcopyright{acmlicensed}
\copyrightyear{2023}
\acmYear{2023}
\acmDOI{10.1145/3588432.3591488}
\acmISBN{979-8-4007-0159-7/23/08}
\acmConference[SIGGRAPH '23 Conference Proceedings]{Special Interest Group on Computer Graphics and Interactive Techniques Conference Conference Proceedings}{August 06--10, 2023}{Los Angeles, CA, USA}
\acmBooktitle{Special Interest Group on Computer Graphics and Interactive Techniques Conference Conference Proceedings (SIGGRAPH '23 Conference Proceedings), August 06--10, 2023, Los Angeles, CA, USA}
\acmPrice{15.00}

%% file: 03-abstract.tex
\begin{abstract}

High-order bases provide major advantages over linear ones in terms of \emph{efficiency}, as they provide (for the same physical model) higher accuracy for the same running time, and \emph{reliability}, as they are less affected by locking artifacts and mesh quality. Thus, we introduce a high-order \ac{FE} formulation (high-order bases) for elastodynamic simulation on high-order (curved) meshes with contact handling based on the recently proposed \ac{IPC} model.

Our approach is based on the observation that each \ac{IPC} optimization step used to minimize the elasticity, contact, and friction potentials leads to linear trajectories even in the presence of nonlinear meshes or nonlinear \ac{FE} bases. It is thus possible to retain the strong non-penetration guarantees and large time steps of the original formulation while benefiting from the high-order bases and high-order geometry.
We accomplish this by mapping displacements and resulting contact forces between a linear collision proxy and the underlying high-order representation.

We demonstrate the effectiveness of our approach in a selection of problems from graphics, computational fabrication, and scientific computing.

\end{abstract}

%% file: 04-teaser.tex
\begin{teaserfigure}
\newcommand{\teaserw}{0.225}
\centering
\includegraphics[width=\linewidth]{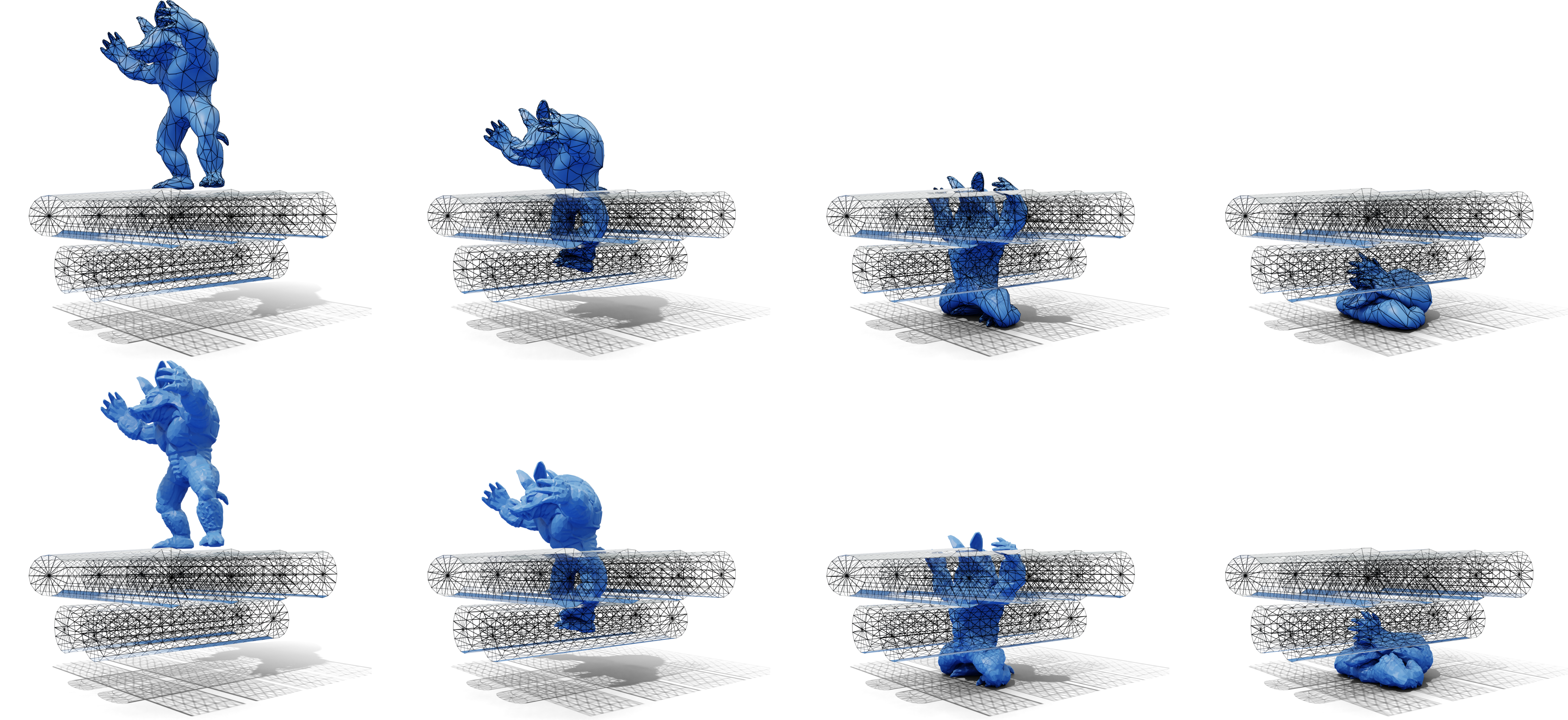}\\
\parbox{\teaserw\linewidth}{\centering $t=\qty{0.0}{\s}$}\hfill
\parbox{\teaserw\linewidth}{\centering $t=\qty{2.5}{\s}$}\hfill
\parbox{\teaserw\linewidth}{\centering $t=\qty{5.0}{\s}$}\hfill
\parbox{\teaserw\linewidth}{\centering $t=\qty{7.5}{\s}$}
\caption[High-order armadillo-rollers]{\figname{High-order armadillo-rollers.} A simulation of an armadillo squished by rollers. We use a high-order volumetric mesh (top row) and deform it with quadratic displacement. To solve collision and compute contact forces, we use a dense linear surface mesh (bottom row) and transfer the deformation and contact forces between the two meshes.}
\label{fig:high-order-teaser}
\end{teaserfigure}

%% file: 10-introduction.tex
\section{Introduction}

Elastodynamic simulation of deformable and rigid objects is used in countless algorithms and applications in graphics, robotics, mechanical engineering, scientific computing, and biomechanics. While the elastodynamic formulations used in these fields are similar, the accuracy requirements differ: while graphics and robotics applications usually favor high efficiency to fit within strict time budgets, other fields require higher accuracy.
In both regimes, \ac{FE} approaches based on a conforming mesh to explicitly partition the object volume are a popular choice due to their maturity, flexibility in handling non-linear material models and contact/friction forces, and convergence guarantees under refinement.

In a \ac{FE} simulation, a set of elements is used to represent the computational domain and a set of basis functions are used within each element to represent the physical quantities of interest (e.g., the displacement in an elastodynamic simulation). Many options exist for both elements and bases. Due to the simplicity of their creation, linear tetrahedral elements are a common choice for the element shape. Similarly, linear Lagrangian functions (often called the hat functions) are often used to represent the displacement field. The linearity in both shape and basis leads to a major and crucial benefit for dynamic simulations: after the displacement is applied to the rest shape, the resulting mesh remains a piece-wise linear mesh. This is an essential property in order to robustly and efficiently detect and resolve collisions~\cite{Wang2021Large}. Collisions between arbitrary curved meshes or between linear meshes over curved trajectories are computationally expensive, especially if done in a conservative way~\cite{Ferguson2021Intersection}.

However, these two choices are restrictive: meshes with curved edges represent shapes, at a given accuracy, with a lower number of elements than linear meshes, especially if tight geometric tolerances are required. Curved meshes are often favored over linear meshes in mechanical engineering~\cite{Hughes2005Isogeometric}. The use of linear bases, especially on simplicial meshes, is problematic as it introduces arbitrary stiffness (a phenomenon known as locking~\cite{Schneider2018Decoupling}). Additionally, high-order bases are more efficient, in the sense that they provide the same accuracy (compared to a reference solution) as linear bases for a lower running time \cite{Babuska1992HP,Schneider2022Large}. Elasto-static problems in computational fabrication (e.g., \cite{Panetta2015Elastic}), mechanics, and biomechanics~\cite{Maas2012Febio} often use high-order bases, but their use for dynamic problems with contact is very limited or the high-order displacements are ignored for contact purposes.

\paragraph{Contribution} We propose a novel elastodynamic formulation supporting both high-order geometry and high-order bases (\cref{fig:high-order-teaser}). Our key observation is that a linear transformation of the displacements degrees of freedom leads to linear trajectories of a carefully designed collision proxy. We use this observation to extend the recently proposed \ac{IPC} formulation, enabling us to use both high-order geometry and high-order bases. Additionally, we can now use arbitrary collision proxies in lieu of the boundary of the \ac{FE} mesh, a feature that is useful, for example, for the simulation of nearly rigid materials.
To evaluate the effectiveness of our approach, we explore its use in graphics applications, where we use the additional flexibility to efficiently simulate complex scenes with a low error tolerance, and we show that our approach can be used to capture complex buckling behaviors with a fraction of the computational cost of traditional approaches. Note that in this work we focus on tetrahedral meshes, but there are no theoretical limitations to applying our method to hexahedral or other polyhedral elements.

\paragraph{Reproducibility} To foster further adoption of our method we release an open-source implementation based on PolyFEM~\cite{PolyFEM} which can be found at \href{https://polyfem.github.io}{polyfem.github.io}.

%% file: 20-related.tex
\section{Related Work}
\label{sec:high-order-related}

\paragraph{High-Order Contacts}

Contact between curved geometries has been investigated in multiple communities, as the benefits of $p$-refinement (i.e., refinement of the basis order) for elasticity have been shown to transfer to problems with contact in cases where an analytic solution is known, such as Hertzian contact \cite{Franke2008P,Franke2010Comparison,Konyukhov2009Incorporation,Aldakheel2020Curvilinear}.

One of the simplest forms of handling contact, penalty methods \cite{Terzopoulos1987Elastically,Moore1988Collision} apply penalty force when objects contact and intersect. However, despite their simplicity and computational advantages, it is well known that the behavior of penalty methods strongly depends on the choice of penalty stiffness (and a global and constant in-time choice ensuring stability may not be possible). \citet{Li2020IPC} propose \ac{IPC} to address these issues, and we choose to use their formulation and benefit from their strong robustness guarantees.

Mortar methods~\cite{Belgacem1998Mortar,Puso2004Mortar,Hueber2006Mortar} are also a popular choice for contact handling, especially in engineering~\cite{Krause2016Parallel} and biomechanics~\cite{Maas2012Febio}. Extensions to high-order \ac{NURBS} surfaces have also been proposed \cite{Seitz2016Isogeometric}.
Mortar methods require to (a priori) mark the contacting surfaces. A clear limitation of this method is that they cannot handle collisions in regions with more than two contacting surfaces or self-collisions. \citet{Li2020IPC} provide a didactic comparison of the \ac{IPC} method and one such mortar method~(\cite{Krause2016Parallel}). They show such methods enforce contact constraints weakly and therefore allow intersections (especially at large timesteps and/or velocities).
Nitsche's method is a method for soft Dirichlet boundary conditions (eliminating the need to tune the penalty stiffness)~\cite{Nitsche1971Uber}. \citet{Stenberg1998Mortaring} and recent work \cite{Gustafsson2020Nitsche, Chouly2022Nitsche} extend Nitsche's method to handle contacts through a penalty or mortaring method. While this eliminates the need to tune penalty stiffnesses, these methods still suffer from the same limitations as mortaring methods.

Another way to overcome the challenges with high-order contact is the use of a \emph{third medium} mesh to fill the empty space between objects~\cite{Wriggers2013Finite}. This mesh is handled as a deformable material with carefully specified material properties and internal forces which act in lieu of the contact forces. In this setting, high-order formulations using $p$-refinement have been shown to be very effective~\cite{Bog2015Normal}. Similar methods have been used in graphics (referred to as an ``air mesh''), as a replacement for traditional collision detection and response methods \cite{Muller2015Air,Jiang2017Simplicial}. The challenge for these approaches is the maintenance of a high-quality tetrahedral mesh in the thin contact regions, a problem that is solved in 2D, but still open for tetrahedral meshes.

The detection and response to collisions between spline surfaces are major open problems in isogeometric analysis, where over a hundred papers have been published on this topic (we refer to \citet{Temizer2011Contact} and \citet{Cardoso2017Contact} for an overview). However, automatic mesh generation for \ac{IGA} is still an open issue \cite{Schneider2021Isogeometric}, limiting the applicability of these methods to simple geometries manually modeled, and often to surface-only problems.

In comparison, we introduce the first technique using the \ac{IPC} formulation to solve elastodynamic problems with contact and friction forces on curved meshes using high-order elements. We also show that an automatic high-order meshing and simulation pipeline is possible when our algorithm is paired with \cite{Jiang2021Bijective}.

\paragraph{High-Order Collision Detection} \Ac{IPC} utilizes \ac{CCD} to ensure that every step taken is intersection-free. The numerical exactness of \ac{CCD} can make or break the guarantees provided by the \ac{IPC} algorithm~\cite{Wang2021Large}. While several authors have proposed methods for collision detection between curved surfaces and nonlinear trajectories \cite{Nelson1998User,VonHerzen1990Geometric,Snyder1993Interval,Nelson2005Haptic,Kry2003Continuous,Ferguson2021Intersection}, there still does not exist a method that is computationally efficient while being conservative (i.e. never misses collisions). Therefore, we are unable to utilize existing methods and instead, propose a method of coupling linear surface representations with curved volumetric geometry.

\paragraph{High-Order Bases}

Linear \ac{FE} bases are overwhelmingly used in graphics applications, as they have the smallest number of \ac{DOF} per element and are simpler to implement. High-order bases have been shown to be beneficial to animate deformable bodies \cite{Bargteil2014Animation}, to accelerate approximate elastic deformations \cite{Mezger2009Interactive}, and to compute displacements for embedded deformations \cite{Longva2020Higher}. Higher-order bases have also been used in meshless methods for improved accuracy and faster convergence~\cite{Martin2010Unified,Faure2011Sparse}.
High-order bases are routinely used in engineering analysis~\cite{Jameson2002Application} where $p$-refinement is often favored over $h$-refinement (i.e., refinement of the number of elements) as it reduces the geometric discretization error faster and using fewer degrees of freedom~\cite{Babuska1988HP,Babuska1992HP,Oden1994Optimal,Bassi1997High,Luo2001Influence}.

We propose a method that allows using high-order bases within the \ac{IPC} framework, thus enabling us to resolve the \ac{IPC} contact model at a higher efficiency for elastodynamic problems with complex geometry, i.e. we can obtain similar accuracy as with linear bases with a lower computation budget. Additionally, our method allows us to explicitly control the accuracy of the collision approximation by changing the collision mesh sampling (\cref{sec:ho-ipc-method}).

High-order bases can be used as a reduced representation and the high-order displacements can be transferred to higher resolution meshes for visualization purposes~\cite{Suwelack2013Accurate}. We use this approach to extend our method to support arbitrary collision proxies, which enables us to utilize our method to accelerate elastodynamic simulations by sacrificing accuracy in the elastic forces.

\paragraph{Physically-Based Simulation}

There is a large literature on the simulation of deformable and rigid bodies in graphics \cite{Bargteil2018Introduction,Kim2022Dynamic}, mechanics, and robotics \cite{Choi2021Use}. In particular, a large emphasis is on the modeling of contact and friction forces \cite{Kikuchi1988Contact,Wriggers1995Finite,Brogliato1999Nonsmooth,Stewart2001Finite}.

\citet{Longva2020Higher} propose a method for embedding geometries in coarser \ac{FE} meshes. By doing so they can reduce the complexity while utilizing higher-order elements to generate accurate elastic deformations. To apply Dirichlet boundary conditions they design the spaces such that they share a common boundary. This scheme, however, cannot capture self-contacts without resorting to using the full mesh. As such they do not consider the handling of contacts. They do, however, suggest a variant of the Mortar method could be future work, but this has known limitations as outlined above. We do not provide a comparison against this method as it does not support contact, and adding contact to it is a major research project on its own, as discussed by the authors.

In our work, we build upon the recently introduced \ac{IPC}~\cite{Li2020IPC} approach, as it offers higher robustness and automation compared to traditional formulations allowing interpenetrations between objects. We review only papers using the \ac{IPC} formulation in this section, and we refer to \cite{Li2020IPC} for a detailed overview of the state of the art.

\citet{Li2020IPC} proposes to use a linear \ac{FE} method to model the elastic potential, and an interior point formulation to ensure that the entire trajectory is free of collisions. While the approach leads to accurate results when dense meshes are used, the computational cost is high, thus stemming a series of works proposing to use reduced models to accelerate the computation. \citet{Li2021Codimensional} propose \ac{CIPC}, a new formulation for codimensional objects is introduced that optionally avoids using volumetric elements to model thin sheets and rod-like objects. An acceleration of multiple orders of magnitude is possible for specific scenes where the majority of objects are codimensional.
\citet{Ferguson2021Intersection} propose a formulation of \ac{IPC} for rigid body dynamics, dramatically reducing the number of \ac{DOF} but adding a major cost and complexity to the collision detection stage, as the trajectories spanned by rigid objects are curved.

\citet{Longva2020Higher} demonstrate their ability to approximately model a rigid body using a single stiff element. This idea is further expanded upon by~\citet{Lan2022Affine} who propose to relax the rigidity assumption: they use an affine transformation to approximate the rigid ones, thus reducing the problem of collision detection to a much more tractable linear \ac{CCD}. Massive speedups are possible for rigid scenes, up to three orders of magnitude compared to the original formulation. While these methods provide major acceleration for specific types of scenes, they are not directly usable for scenes with deformable objects.

\citet{Lan2021Medial} proposes to use medial elastics \cite{Lan2020Medial}, a family of reduced models tailored for real-time graphics applications. In their work, the shape is approximated by a medial skeleton which is used to model both the elastic behavior and as a proxy for collision detection. The approach can simulate deformable objects, however, it cannot reproduce a given polyhedral mesh and it is also specialized for medial elasticity simulations.

In our work, we enable the use of high-order meshes and high-order elements in a standard \ac{FE} framework. Our approach decouples the mesh used to model the elastic potential from the mesh used for the contact and friction potentials, thus providing finer-grained control between efficiency and accuracy. %

\paragraph{Convergence and use of $C^0$ Lagrangian Elements}
Studies compare $C^0$ (p-\ac{FEM}) and \ac{IGA} bases' convergence under p-refinement \cite{Sevilla2011Comparison}, in the presence of contact \cite{Temizer2011Contact,Seitz2016Isogeometric} and in other settings such as electromechanics \cite{Poya2018Curvilinear}. \ac{IGA} bases have been shown, in specific problems with simple geometries, to have slightly higher accuracy compared to Lagrangian $C^0$ elements. In this work, we favor Lagrangian $C^0$ elements as \ac{IGA} meshes are hard to generate for complex geometries and, additionally, some of their benefits are lost when non-regular grid meshes are required to represent complex geometry \cite{Schneider2022Large,Schneider2019Poly}. Our paper does not study the convergence of the method, we leave a convergence (h and p) study as future work jointly with a convergence study for the \ac{IPC} contact model. Our goal is restricted to show that elastodynamic simulations with high-order geometry and bases are possible on complex geometry and provide a practical speedup over the linear geometry representation and linear bases that are commonly used in graphics applications.

%% file: 30-background.tex
\section{IPC Overview}\label{sec:ipcsummary-ho}

Our approach builds upon the \ac{IPC} solver introduced in
\cite{Li2020IPC}. In this section, we review the original formulation and introduce the notation.

\citet{Li2020IPC} computes the updated displacements $u^{t+1}$ of the objects at the next time step by solving an \emph{unconstrained} non-linear energy minimization:
\begin{equation}
  u^{t+1} = \argmin_u \> E(u, u^t, v^t) + B(x+u,\hat{d}) + D(x+u, \epsilon_v),
  \label{eq:ipc-in-ho}
 \end{equation}
where $x$ is the vertex coordinates of the rest position, $u^t$ is the displacement at the current step, $v^t$ the velocities, $E(u, u^t, v^t)$ is a time-stepping \ac{IP}~\cite{Kane2000Variational}, $B$ is the barrier potential, and $D$ is the lagged dissipative potential for friction~\cite{Li2020IPC}. The user-defined geometric accuracy $\hat{d}$ controls the maximal distance at which the barrier potential will have an effect. Similarly, the smooth friction parameter $\epsilon_v$ controls the smooth transition between static and dynamic friction. We refer to \citet{Li2020IPC} for a complete description of the potentials, as for our purposes we will not need to modify them.

\paragraph{Solver and Line Search CCD}
The advantage of the \ac{IPC} formulation is that it is possible to prevent intersections from happening by using a custom Newton solver with a line-search that explicitly checks for collisions using a continuous collision detection algorithm \cite{Provot1997Collision,Wang2021Large}, while keeping the overall simulation cost comparable to the more established \ac{LCP} based contact solvers \cite{Li2020IPC}.

%% file: 40-method.tex
\begin{figure}[t]
    \centering
    \begin{minipage}{0.3\linewidth}
        \includegraphics[width=\linewidth]{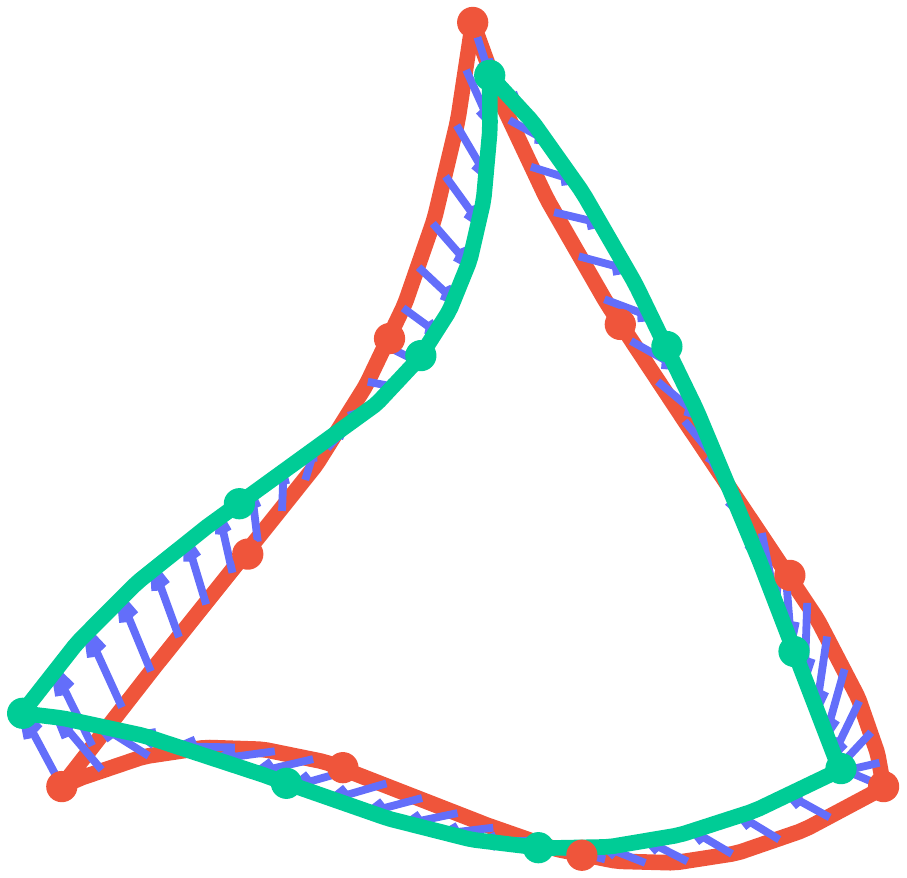}
    \end{minipage}%
    \begin{minipage}{0.7\linewidth}
        \centering
        \begin{align*}
            &\left(\sum_{i=1}^n u_i^{t+1} \basis_i \right)
            - \left(\sum_{i=1}^n u_i^{t} \basis_i\right) \\
            &= \sum_{i=1}^n (u_i^{t+1} - u_i^{t}) \basis_i = \sum_{i=1}^n \Delta u_i \basis_i
        \end{align*}
    \end{minipage}
    \caption[Linearity of displacement update]{\figname{Linearity of displacement update.} Even with nonlinear bases $\basis_i$, the update to displacement still constitutes a linear combination of nodal displacements. Therefore from a starting position (in red), the update to displacements of any point on the surface (in blue) is linear, and as such we need not use expensive nonlinear CCD.}
    \label{fig:linear_disp}
    \vspace{-2mm}
\end{figure}

\section{Method}\label{sec:ho-ipc-method}

We introduce an extension of \ac{IPC} for a curved mesh $\M=(V_{\M}, T_{\M})$
where $V_{\M}$ and $T_{\M}$ are the nodes and volumetric elements of $\M$, respectivly.
The formulation reduces to standard \ac{IPC} when linear meshes and linear bases are used, but other combinations are also possible: for example, it is possible to use high-order bases on standard piece-wise linear meshes, as we demonstrate in \cref{sec:results}.

We first introduce explicit definitions for functions defined on the volume and the contact surface corresponding to its boundary.
Let $f_\M \colon \M \to \real^3$ be a volumetric function (in our case the volumetric displacement $u$) defined as
\begin{equation}
f_\M = \sum_{i=1}^n f_\M^i \basis_\M^i,
\label{eq:volume}
\end{equation}
where $\basis_\M^i$ are the $n$ \ac{FE} bases defined on $\M$ and $f_\M^{i}$ their coefficient.

Similarly on the surface $\S=(V_{\S}, T_{\S})$ used for collision, with vertices $V_{\S}$ and triangular faces $T_{\S}$, we define $f_\S \colon \S \to \real^3$ (in our case the displacement $u$ restricted to the surface) as
\begin{equation}
f_\S = \sum_{j=1}^m f_\S^j \basis_{\S}^{j},
\label{eq:surface}
\end{equation}
where $\basis_\S^j$ are the $m$ \ac{FE} bases defined on $\S$ and $f_\S^{j}$ their coefficient. We can now rewrite \cref{eq:ipc-in-ho} to make explicit that the potential $E$ depends on $\M$, while $B$ and $D$ only depend on $\S$:
\begin{equation}
\begin{split}
  u^{t+1} = \argmin_u \> &E_{\M}(u, u^t, v^t) + B_{\S}(V_{\S}+\Phi(u),\hat{d}) \\
  &+ D_{\S}(V_{\S}+\Phi(u), \epsilon_v),
\end{split}
\label{eq:formulation}
\end{equation}
where $\Phi \colon \vecspace_{\M} \to \vecspace_{\S}$ is an operator where $\vecspace_{\M} = \sspan{\basis_\M^i}$ and $\vecspace_{\S} = \sspan{\basis_\S^j}$ that transfers volumetric functions on $\M$ to $\S$. In the context of \cite{Li2020IPC} (i.e., \cref{eq:ipc-in-ho}), $\Phi$  is a restriction of the volumetric function to its surface. While in general, $\Phi$ could be an arbitrary operator, \ac{IPC} takes advantage of its linearity: if $\Phi$ is linear, then the trajectories of surface vertices in one optimization step of \cref{eq:formulation} will be linear (\cref{fig:linear_disp}), and it is thus possible to use standard continuous collision detection methods~\cite{Provot1997Collision,Wang2021Large}. If $\Phi$ is nonlinear, for example in the rigid-body formulation introduced by \citet{Ferguson2021Intersection}, the collision detection becomes considerably more expensive~\cite{Lan2022Affine}.

We observe that arbitrary \emph{linear} operators can be used for $\Phi$, and note that increasing the order of the bases used to represent $f_\M$ and $f_\S$ does not affect the linearity of the operator. An additional advantage of this reformulation is that the space $\vecspace_{\S}$ does not have to be a subspace of $\vecspace_{\M}$. For example, the collision mesh can be at a much higher resolution than the volumetric mesh used to resolve the elastic forces (\cref{sec:results}). %

We first discuss how to build a linear operator $\Phi$ for high-order meshes, high-order elements, and arbitrary collision proxies, and we postpone the discussion on how to adapt the \ac{IPC} algorithm to work with arbitrary $\Phi$ to \cref{sec:grad}.

\subsection[Construction of the transfer operator]{Construction of $\Phi$}

We present two methods for constructing $\Phi$: upsampling the surface of $\M$ to obtain a dense piecewise linear approximation of its boundary, which we use as $\S$ (\cref{sec:upsample}), or using an arbitrary surface triangle mesh as $\S$ and determining closest point correspondences used to evaluate bases (\cref{sec:arbitrary-proxy}). Our results in \cref{sec:results} show a mix of both approaches: \cref{fig:beam-bending,fig:bouncing-ball,fig:mat-twist,fig:microstructure,fig:arma-roller-proxy} use an upsampling while \cref{fig:high-order-teaser,fig:bouncing-ball,fig:arma-balance,fig:arma-roller-proxy,fig:trash-compactor,fig:screw,fig:rolling_ball_friction} use an arbitrary triangle mesh proxy.

Since $\Phi$ is a linear operator, a discrete function $f_\M \in \vecspace_{\M}$ with coefficients $f_\M^{i}$ can be transferred to $f_\S \in \vecspace_{\S}$ using its $m$ coefficients $f_\S^j$ as
\[
\vec{f}_\S = \W \vec{f}_\M,
\]
where $\vec{f}_\M$ and $\vec{f}_\S$ are the stacked coefficients $f_\M^i$ and $f_\S^j$, respectively.
The tetrahedron $t^i_{\M}\in T_{\M}$ of a high-order mesh $\M$ is defined as the image of the \emph{geometric mapping} $g^i$
applied to reference right-angle tetrahedron $\hat t$; that is
\[
t^i_{\M} = g^i(\hat t).
\]
On $\S$, the geometric map is a vectorial function and has the same form as \cref{eq:surface}.

\subsubsection{Upsampled linear boundary}\label{sec:upsample}

To construct $\S$ we need to use the \emph{geometric map} to find the initial vertex positions, while to define the operator to transfer functions from the volumetric mesh to $\S$ we will use the \emph{basis functions} of $\M$.

\paragraph{Vertex Positions}
Every vertex of the piece-wise linear approximation $v^j_{\S}\in V_{\S}$ has coordinates $\hat v^j$ in the reference tetrahedron of $t^i_{\M}$, so its global coordinates can be computed as
\[
v^j_{\S} = g^i(\hat v^j),%
\]
and stacked into the vector $V_{\S}$ used in \cref{eq:formulation}.

\paragraph{Transfer}

To construct the linear operator $\Phi$ encoded with the matrix $\W$ transferring from a higher-order polynomial basis on the boundary of $\M$ to the piecewise linear approximation $\S$, we observe that, since $\S$ is an upsampling of $\M$, we can use $\hat v^j$ to directly evaluate the bases of $\M$ (for all non-zero bases) and use them as a weight to transfer the function from $\M$ to $\S$ and define
\[
\W_{ji} = \basis_{\M}^i(\hat v^j),
\]
which is a linear operator, independent of the degree of the basis functions.

\subsubsection{Arbitrary Triangle Mesh Proxy}\label{sec:arbitrary-proxy}

\begin{algorithm}[t]
\centering
\caption{Construct $\Phi=W\vec{f}_\M$ for arbitrary triangle mesh}
\label{alg:arbitrary-proxy}
\begin{algorithmic}[1]
\STATE $W\gets \bf{0}$ \COMMENT{$W \in \real^{{m}\times{n}}$}
\STATE $\linearM \gets \operatorname{linearize}(\M, 4)$ \COMMENT{4 linear tetrahedra per curved tet}
\FOR{$v^j_\S \in V_\S$}
\STATE $\aabb \gets \operatorname{inflate}(\operatorname{AABB}(v^j_\S), 10^{-3})$
\WHILE[$n=3$ in our examples]{$|\aabb \cap \linearM| < n$}
    \STATE $\aabb \gets \operatorname{inflate}(\aabb, 10\%)$
\ENDWHILE
\FOR{{$t^i_\M \in (\aabb \cap \M)$}}
    \STATE $\tilde{b}_i \gets \operatorname{BC}(v^j_\S, \operatorname{linearize}(t^i_\M))$
    \COMMENT{barycentric coords.}
    \STATE $\hat{v}^j_i \gets \argmin_{\hat v} \> \| g^i(\hat v) - v^j_{\S}\|^2_2$
    \COMMENT{\acs{LBFGS} with $\hat{v}_0=\tilde{b}_i$}
\ENDFOR
\STATE $i^* \gets \argmin_i \> \|\hat v^j_i\|_1$
\COMMENT{Closest to the interior}
\STATE $\hat v^j = \hat{v}^j_{i^*}$ \COMMENT{pre-image of $v^j_{\S}$}
\STATE $W_{ji} = \basis_\M^i(\hat v^j)$
\ENDFOR
\RETURN $W$
\end{algorithmic}
\end{algorithm}
\vspace{-2mm}

The same construction applies to arbitrary mesh proxies (e.g., \cref{fig:high-order-teaser}), but we need to compute $\hat v^j$ for every vertex. When $\M$ is linear we can simply compute $\hat v^j$ as the barycentric coordinates of the closest tetrahedron in $\M$, but when $\M$ is nonlinear we use an optimization to invert $g^i$~\cite{Suwelack2013Accurate}. However, unlike \citet{Suwelack2013Accurate}, we found that using a normal field to define correspondences is fragile when the surfaces have a very different geometric shape, so we opt for a simpler formulation based on distances.

\Cref{alg:arbitrary-proxy} outlines our method for computing $\Phi$ for an arbitrary triangle proxy. Namely, given a volumetric mesh $\M$ and an arbitrary triangle mesh $\S$ we do not have the pre-image under the geometric mapping of the vertices $v^j_{\S}\in V_{\S}$, so we compute one by determining the closest element in $\M$ to $v^j_{\S}$ and use an optimization to compute the inverse geometric mapping to obtain the coordinates $\hat{v}^j$.
This procedure only needs to be performed once because $W$ depends only on the rest geometry.

\subsection{Gradient and Hessian of Surface Terms}\label{sec:grad}

Adapting \ac{IPC} to work with arbitrary linear $\Phi$ mapping requires only changing the \emph{assembly} phase, which requires gradients and Hessian of the surface potentials. Similar to \ac{IPC}, we use Newton's method to minimize the newly formulated potential in \cref{eq:formulation}, and we thus need its gradient and Hessian.

For a surface potential $B_{\S}(V_{\S}+\Phi(u),\hat{d})$ and transfer
\[
\Phi(u) = \Phi\left(\sum_{i=1}^n u_i \basis_\M^i\right) = \sum^m_{j=1} (\W \vec{u})_j \basis_\S^j,
\]
where $\vec{u}$ is the \emph{vector} containing all the coefficients $u_i$; we use the definition of $\W$ to express the gradient of the barrier (or the friction) potential as
\[
\begin{split}
\nabla_{u}& B_{\S}(V_{\S}+\Phi(u),\hat{d}) = \nabla_{u} \transpose{(V_{\S} + \Phi(u)))} \nabla_{\S_u}B_{\S}(\S_u,\hat{d})\\
&= \nabla_{u} (V_{\S} + \transpose{(\W \vec{u}))} \nabla_{\S_u}B_{\S}(\S_u,\hat{d}) = \transpose{\W} \nabla_{\S_u}B_{\S}(\S_u,\hat{d}),
\end{split}
\]
where $\S_u = V_{\S} + \Phi(u)$. The Hessian is computed similarly
\[
\nabla^2_{u} B_{\S}(V_{\S}+\Phi(u),\hat{d}) = \transpose{\W} [\nabla^2_{\S_u}B_{\S}(\S_u,\hat{d})] \W.
\]
The formulas for $\nabla_{\S_u}B_{\S}(\S_u,\hat{d})$, $\nabla_{\S_u}D_{\S}(\S_u, \epsilon_v)$, and their Hessians are the same as in \cite{Li2020IPC}, thus requiring minimal modifications to an existing implementation.
As in \cite{Li2020IPC}, we mollify the edge-edge distance computation to avoid numerical issues with nearly parallel edges.

%% file: 50-results.tex
\section{Results}\label{sec:results}

\input{figs/beam-bending}

All experiments are run on individual nodes of an \ac{HPC} cluster each using two Intel Xeon Platinum 8268 24C 205W \qty{2.9}{\GHz} Processors and 16 threads. All results are generated using the PolyFEM~library \cite{PolyFEM} coupled with the IPC~Toolkit~\cite{IPCToolkit}, and use the direct linear solver Pardiso~\cite{Alappat2020Recursive,Bollhofer2020State,Bollhofer2019Large}. We use the notation $P_n$ to define the \ac{FE} bases order (e.g., $P_2$ indicates quadratic Lagrange bases) and all our curved meshes are quartic. All simulation parameters and a summary of the results can be found in \cref{tab:sim_params,tab:sim_results}.

\subsection{Test Cases}\label{sec:test-cases}

\paragraph{Bending beam} We first showcase the advantages of high-order bases and meshes. \Cref{fig:beam-bending} shows that linear bases on a coarse mesh introduce artificial stiffness and the result is far from the reference (a dense $P_1$ mesh). As we increase the order, the beam bends more. Using $P_3$ on such a coarse mesh leads to results indistinguishable from the reference at a fraction of the cost. We also compare the results of a higher resolution $P_1$ mesh with a limited time budget. That is, the number of elements is chosen to produce a similar running time as the $P_3$ results (1,124 tetrahedra compared to 52 in the coarse version). Even in this case, the differences are obvious and far from the expected results. %

\input{figs/bouncing-ball}

\paragraph{Bouncing ball} \Cref{fig:bouncing-ball} shows the movement of the barycenter of a coarse bouncing sphere on a plane. When using linear bases on the coarse mesh, the ball tips over and starts rolling as the geometry is poorly approximated (yellow line). Replacing the coarse collision mesh using our method (blue line) improves the results for a small cost (\qty{125}{frames\per\s} versus \qty{83.3}{frames\per\s}); however, since the sphere boundary is poorly approximated and the bases are linear, the results are still far from the accurate trajectory (green line). Finally, replacing $\M$ with a curved mesh and using $P_2$ bases leads almost to the correct dynamics (red line) while maintaining a real-time simulation (\qty{38.4}{frames\per\s}). As a reference, the dense $P_1$ linear mesh (green line) runs at \qty{3.9}{frames\per\s}.

\paragraph{Rolling ball}
\Cref{fig:rolling_ball_friction} shows our method is able to maintain purely tangential friction forces on the \ac{FE} mesh while rolling a ball down a slope. The baseline spherical \ac{FE} mesh (8.8K $P_1$ tetrahedra) and our method using a cube \ac{FE} mesh (26 $P_1$ tetrahedra), both using the same collision geometry, produce very similar dynamics, but our method is $7.5\times$ faster. However, while the ball's material is stiff ($E=10^9~\unit{\Pa}$), it is not rigid, so the baseline model deforms slightly at the point of contact. Our model exhibits extra numerical stiffness from the large linear elements and so deforms less. This results in a 5\% difference on average in the minimum distance which translates to a normal force (and ultimately friction force) that is $2\times$ greater. This inaccuracy is a limitation of using such a course \ac{FE} mesh within our framework.

\subsection{Examples}

\input{figs/mat-twist.tex}

\paragraph{Mat twist} We reproduce the mat twist example in \cite{Li2020IPC} using a thin linear mesh $\M$ with 2K tetrahedra and simulate the self-collisions arising from rotating the two sides using a collision mesh $\S$ with 65K vertices (\cref{fig:mat-twist}). Simulating this result using standard \ac{IPC} on the coarse (left) is fast but leads to visible artifacts; by using $P_2$ bases for displacements the results are smooth and the simulation is faster (\qty{91}{\s\per{frame}}). For reference, a finer linear solution with more elements, to get a result similar to ours but only using linear elements, requires 230K elements and a runtime 10$\times$ higher.

We find a $P_1$ mesh with 51K tetrahedra ($25\times$ the number used in the $P_2$ variant) that produces a similar running time. The $P_2$ collision mesh uses $3.5\times$ more triangles leading to $3.1\times$ slower collision detection while the linear solver for the $P_1$ mesh is only $2.2\times$ slower. This results in similar dynamics and final state (see \cref{fig:mat-twist}) with some notable differences around the folds of the mat.

\paragraph{Microstructure} In \cref{fig:microstructure}, we simulate the compression of an extremely coarse (6K $P_4$ tetrahedra) curved microstructure mesh from~\cite{Jiang2021Bijective}. We upsample its surface to generate a collision mesh with 143K triangles. We demonstrate our method's ability to simulate \emph{anisoparametric} scenarios (i.e., the shape and basis functions differ) by using $P_1$ and $P_2$ bases. In this case, both simulations take a similar amount of time (\hms{6}{34}{9} versus \hms{6}{4}{48}).

\paragraph{Armadillo on a Roller} In \cref{fig:arma-roller-proxy}, we replicate the armadillo roller from~\cite{Verschoor2019Efficient} and use fTetWild~\cite{Hu2020Fast} to generate $\M$ with 1.8K tetrahedra (original mesh has 386K). With our method, we combine $\M$ with the original surface with 21K faces with linear element and obtain a speedup of 60$\times$ (row$^\star$). We used~\cite{Jiang2021Bijective} to generate a coarse curved mesh (with only 4.7K tetrahedra) and use an optimization to invert the geometric mapping and simulate the result using $P_2$, this leads to a simulation 30$\times$ faster (row$^\dagger$). Finally, we upsampled the surface of the curved mesh to generate a new collision mesh $\S$ with 20K faces, this simulation is only 8$\times$ faster (row$^\ddagger$).

\paragraph{Trash-compactor}  We reproduce the trash compactor from \cite{Li2020IPC} using a coarse mesh $\M$ with 21K tetrahedra and compress it with five planes. Since the input mesh is already coarse and the models have thin features in the tentacles, we use fTetWild to generate a coarser mesh with 3.5K tetrahedra. Using this mesh with $P_1$ displacements while using the same surface mesh for collisions provides a 2.5$\times$ speedup. Since both coarse and input meshes have similar resolution, using $P_2$ leads to a more accurate but much slower (around 10$\times$) result as the number of \ac{DOF} for $P_2$ is similar to the denser mesh but with 5$\times$ the number of surface triangles.

\subsection{Extreme coarsening}

\paragraph{Nut and Bolt} As mentioned in \cref{sec:ho-ipc-method}, our method can be used with linear meshes and linear bases. This is best suited to stiff objects where the deformation is minimal. \Cref{fig:screw} shows an example of a nut sliding inside a bolt, since both materials are stiff ($E=\qty{200}{\GPa}$), we coarsen $\M$ using fTetWild~\cite{Hu2020Fast} from 6K tetrahedra and 1.7K vertices to 492 and 186, respectively. This change allows our method to be twice as fast without visible differences.

\input{figs/arma-balance}

\paragraph{Balancing Armadillo} When generating a coarse mesh $\M$ the center of mass and mass of the object might change dramatically. \Cref{fig:arma-balance} shows that the coarse mesh cannot balance anymore as the center of mass is outside the contact area. To prevent this artifact, similarly to~\cite{Prevost2013Make}, we modify the density (in red in the third figure) of the material to move the center of mass.

%% file: figs/beam-bending.tex
\begin{figure}[t]
    \centering\footnotesize
    \includegraphics[width=0.95\linewidth]{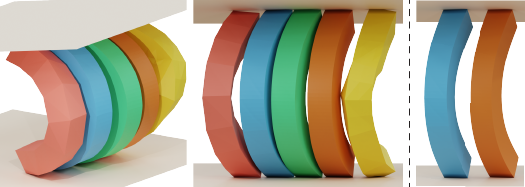}
    \begin{tabular}{ccccc}
        {\color[HTML]{EC7063}$P_1$ coarse} &
        {\color[HTML]{5DADE2}$P_1$ reference} &
        {\color[HTML]{58D68D}$P_2$} &
        {\color[HTML]{DC7633}$P_3$} &
        {\color[HTML]{c19c0b}$P_1$ time budgeted} \\
        {\color[HTML]{EC7063}(\qty{15}{\s})} &
        {\color[HTML]{5DADE2}(\ms{7}{43})} &
        {\color[HTML]{58D68D}(\qty{32}{\s})} &
        {\color[HTML]{DC7633}(\qty{58}{\s})} &
        {\color[HTML]{c19c0b}(\qty{57}{\s})}
        \vspace{-2mm}
    \end{tabular}
    \caption[Bending beam]{\figname{Bending beam.} Squared-section coarse beam pressed by two planes. Linear elements exhibit artificial stiffness as they cannot bend. The reference $P_1$ solution and $P_3$ are rendered in isolation on the right. The results are indistinguishable, but $P_3$ is an order of magnitude faster.}
    \label{fig:beam-bending}
    \vspace{-2mm}
\end{figure}

%% file: figs/bouncing-ball.tex
\begin{figure}[t]
    \centering\footnotesize
    \parbox{.06\linewidth}{~}\hfill
    \parbox{.3\linewidth}{\centering $t=0.15$}\hfill
    \parbox{.3\linewidth}{\centering $t=0.25$}\hfill
    \parbox{.3\linewidth}{\centering $t=1.0$}\\

    \parbox{.06\linewidth}{\centering\rotatebox[origin=c]{90}{\parbox{2.5\linewidth}{\centering Combined}}}\hfill
    \parbox{.3\linewidth}{\includegraphics[width=\linewidth]{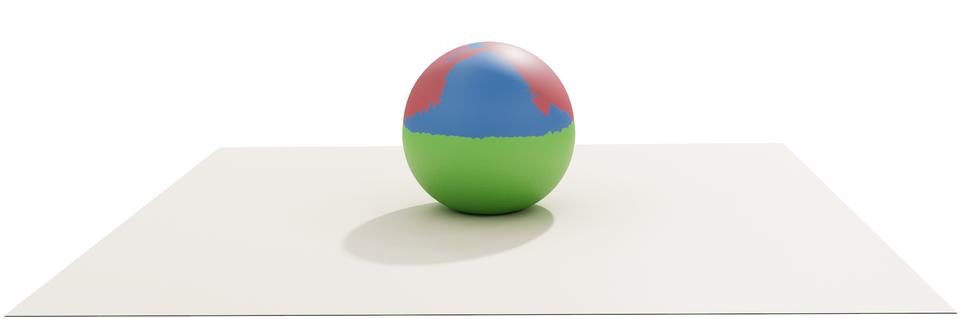}}\hfill
    \parbox{.3\linewidth}{\includegraphics[width=\linewidth]{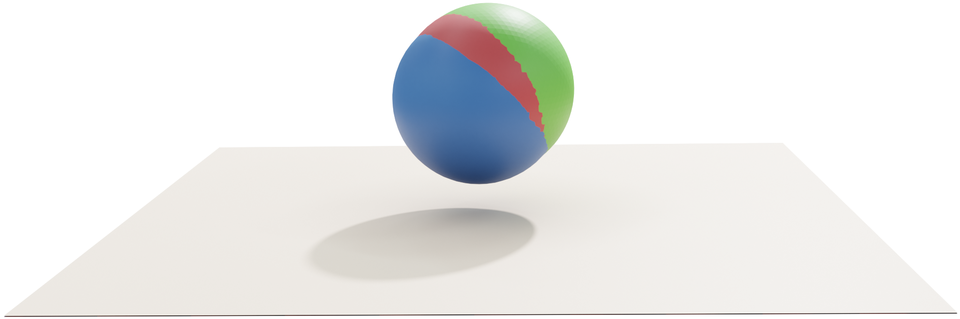}}\hfill
    \parbox{.3\linewidth}{\includegraphics[width=\linewidth]{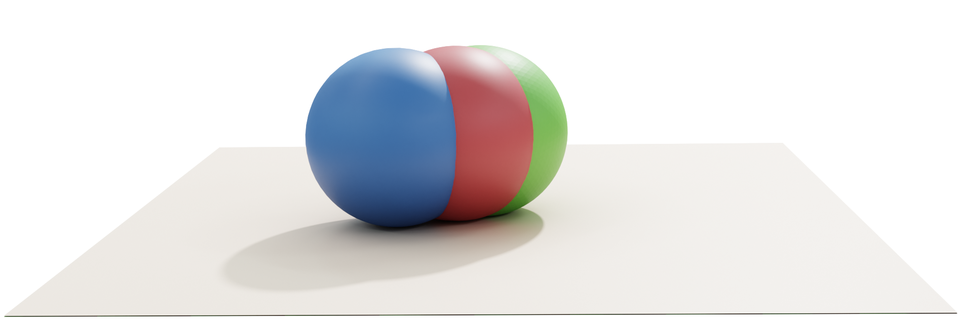}}\\

    \parbox{.06\linewidth}{\centering\rotatebox[origin=c]{90}{\parbox{2.5\linewidth}{\centering Coarse}}}\hfill
    \parbox{.3\linewidth}{\includegraphics[width=\linewidth]{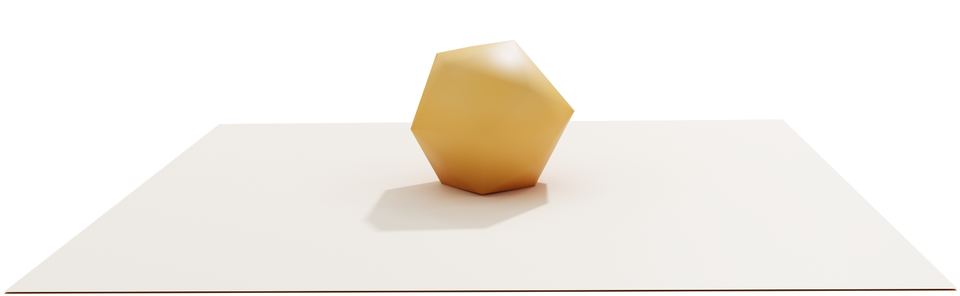}}\hfill
    \parbox{.3\linewidth}{\includegraphics[width=\linewidth]{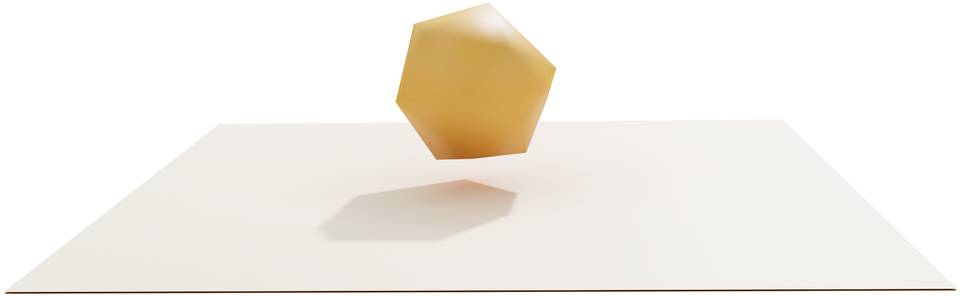}}\hfill
    \parbox{.3\linewidth}{\includegraphics[width=\linewidth]{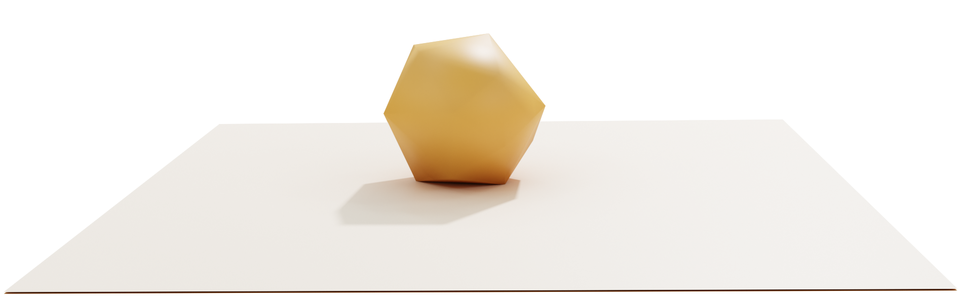}}\\

    \parbox{.06\linewidth}{\centering\rotatebox[origin=c]{90}{\parbox{2.5\linewidth}{\centering Coarse $P_4$\\FE Mesh}}}
    \parbox{.3\linewidth}{\includegraphics[width=\linewidth]{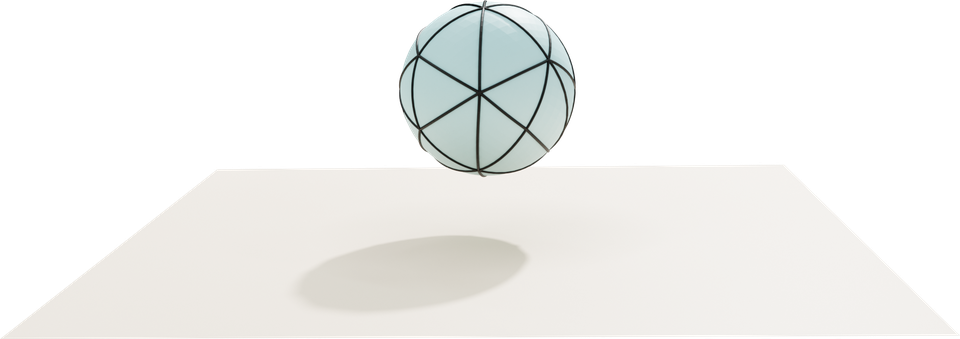}}

    \includegraphics[width=0.9\linewidth]{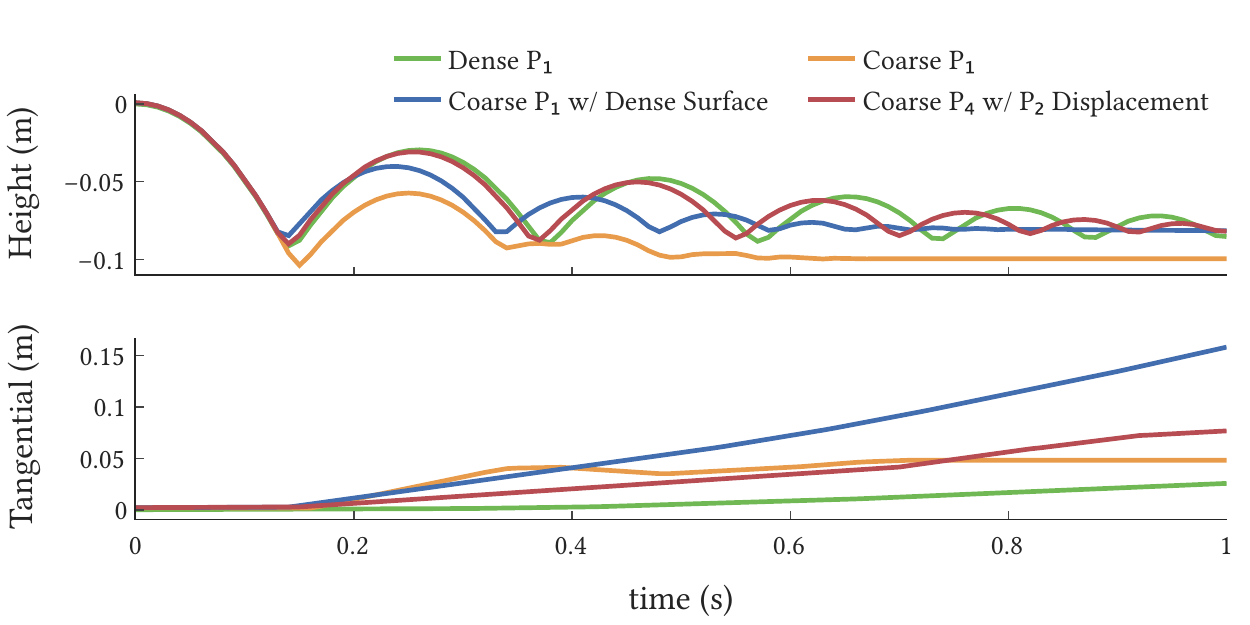}

    \caption[Bouncing ball]{\figname{Bouncing ball.} Simulation of a bouncing sphere on a plane. The yellow image and line are the baseline, a coarse linear mesh with linear displacement. The results can be improved using our method and replacing $\S$ with a dense sphere in blue. When using a high-order mesh with $P_2$ displacement, red, the results are similar to the dense linear simulation in green.}
    \label{fig:bouncing-ball}
    \vspace{-2mm}
\end{figure}

%% file: figs/mat-twist.tex
\begin{figure}[!t]
    \centering\footnotesize
    \setlength\tabcolsep{0pt}
    \begin{tabular}{cccc}
        \makecell{$P_1$ coarse\\(\ms{2}{47})} &
        \makecell{$P_1$ time budgeted\\(\hms{6}{7}{12})} &
        \makecell{$P_2$\\(\hms{6}{19}{52})} &
        \makecell{$P_1$\\(\dhms{2}{14}{13}{0})}\\
        \includegraphics[width=.249\linewidth]{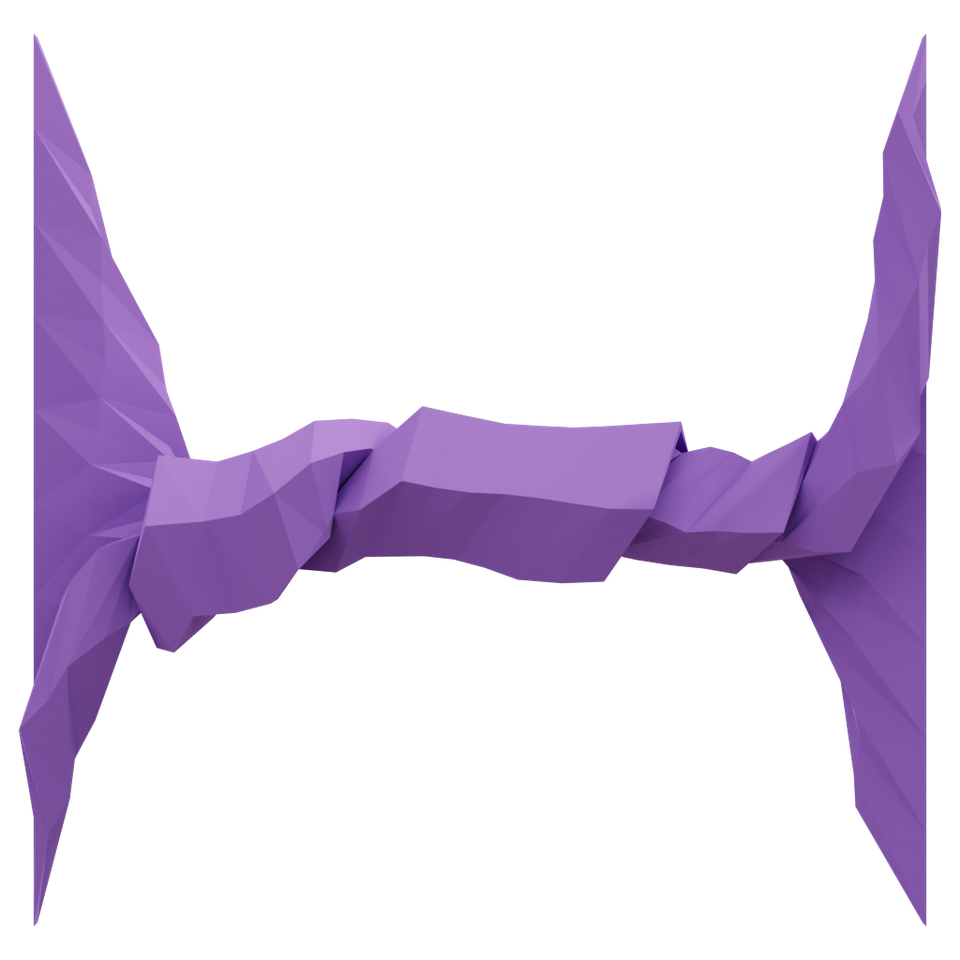} &
        \includegraphics[width=.249\linewidth]{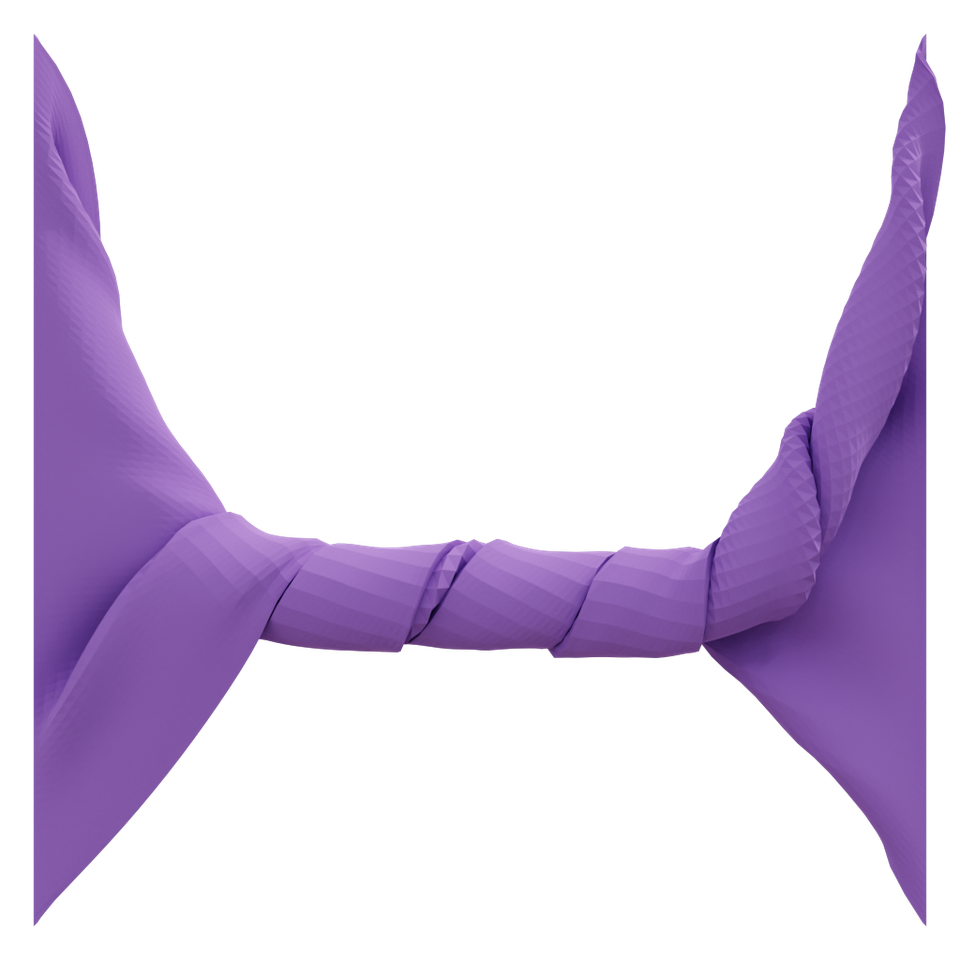} &
        \includegraphics[width=.249\linewidth]{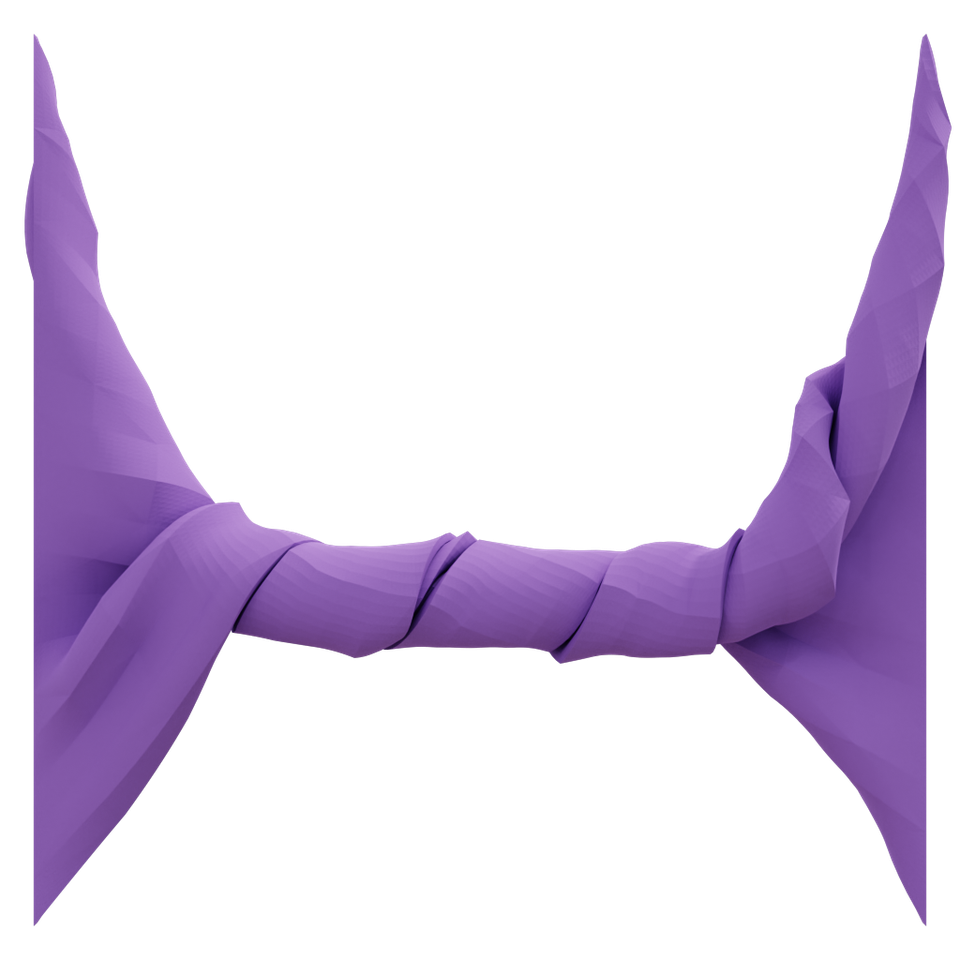} &
        \includegraphics[width=.249\linewidth]{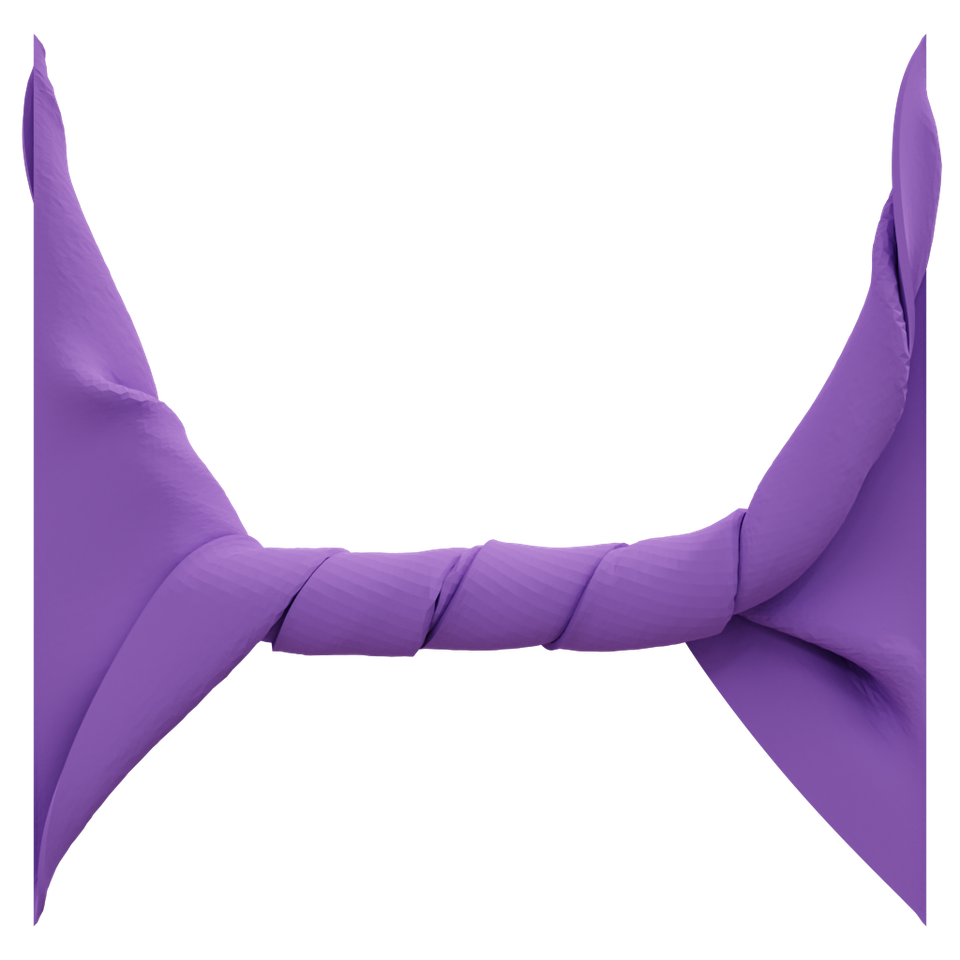}\\[-2.75mm]
        \includegraphics[width=.249\linewidth]{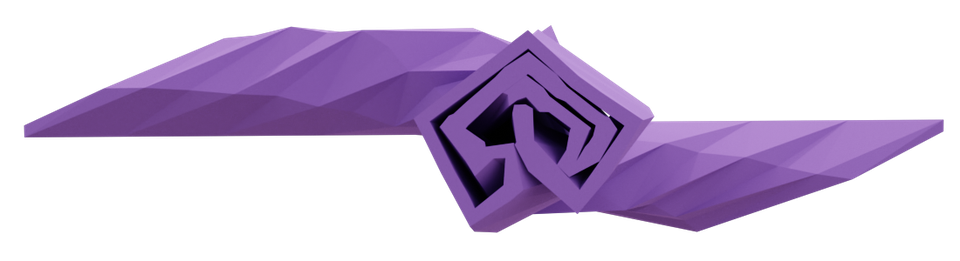} &
        \includegraphics[width=.249\linewidth]{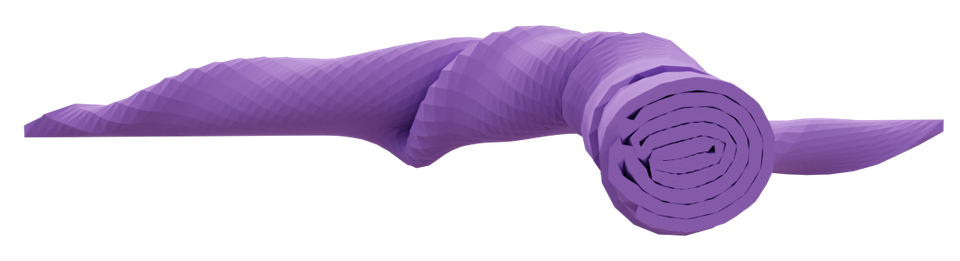} &
        \includegraphics[width=.249\linewidth]{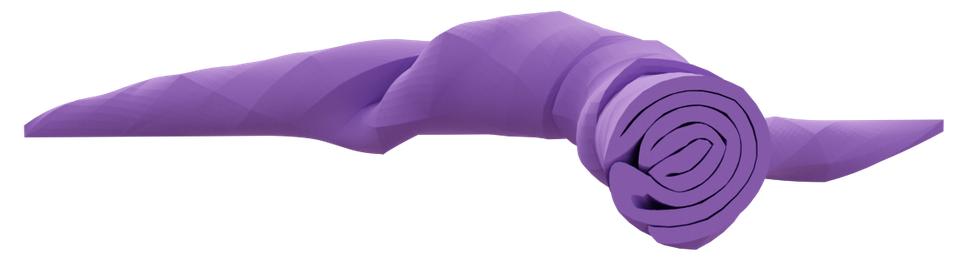} &
        \includegraphics[width=.249\linewidth]{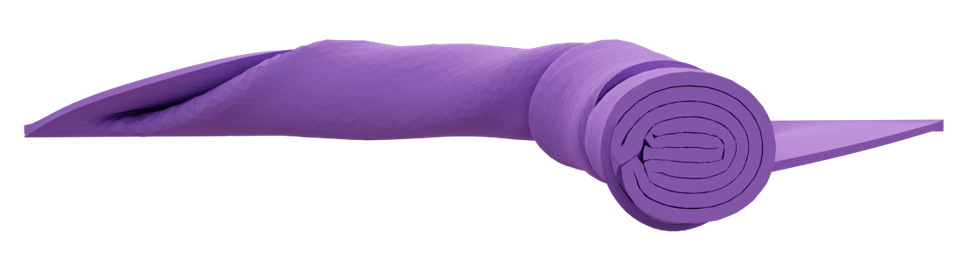}
    \end{tabular}
    \vspace{-4mm}
    \caption[Mat-twist]{\figname{Mat-twist.} Simulation of twisting for different bases' order and mesh resolutions. The cross-section (bottom row) shows that the coarse linear mesh (left) has huge artifacts. The coarse $P_2$ bases (middle-right) produce smooth results similar to a dense mesh (right) for a tenth of the time. A ``time-budgeted'' version shows similar results but exhibits checker patterns around the folds.}
    \label{fig:mat-twist}
    \vspace{-3mm}
\end{figure}

%% file: figs/arma-balance.tex
\begin{figure}[!t]
    \centering\footnotesize
    \setlength{\tabcolsep}{.01\linewidth}
    \begin{tabular}{c|ccc}
        Initial config. &
        Fine (17s) & Coarse (19s) & Optimized (9s)\\\\
        \includegraphics[width=.22\linewidth]{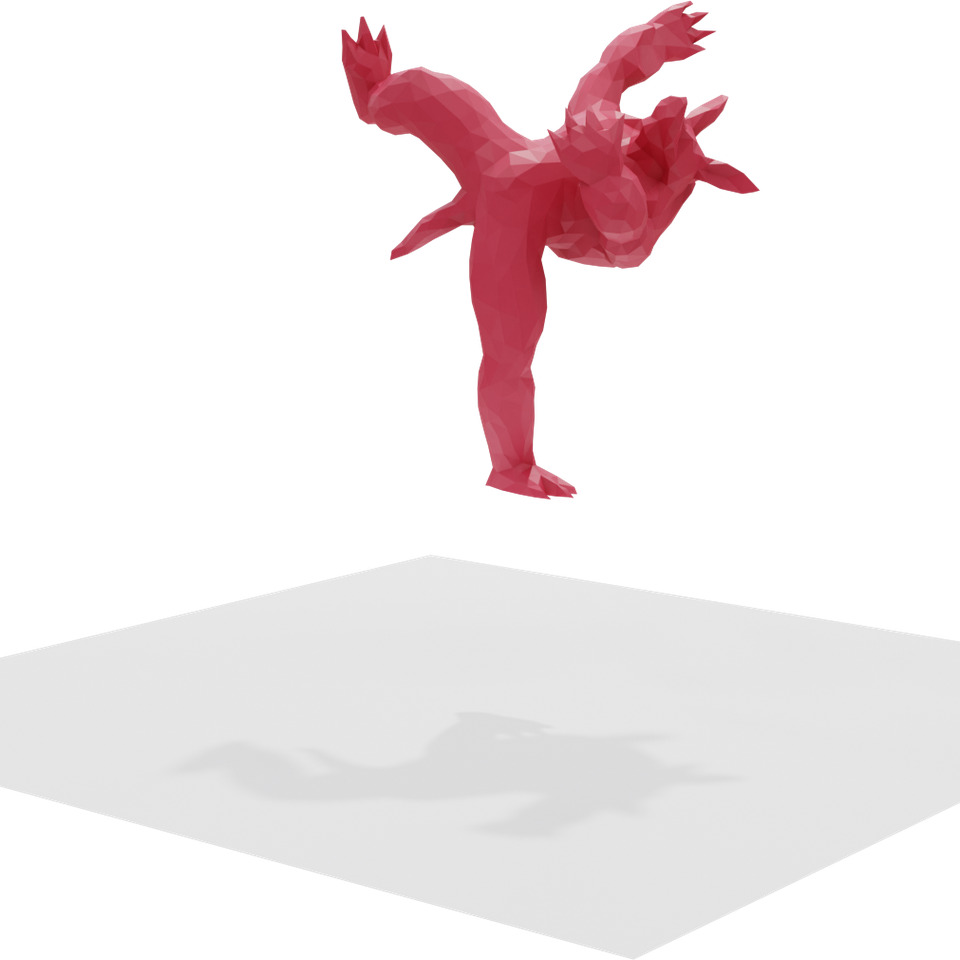} &
        \includegraphics[width=.22\linewidth]{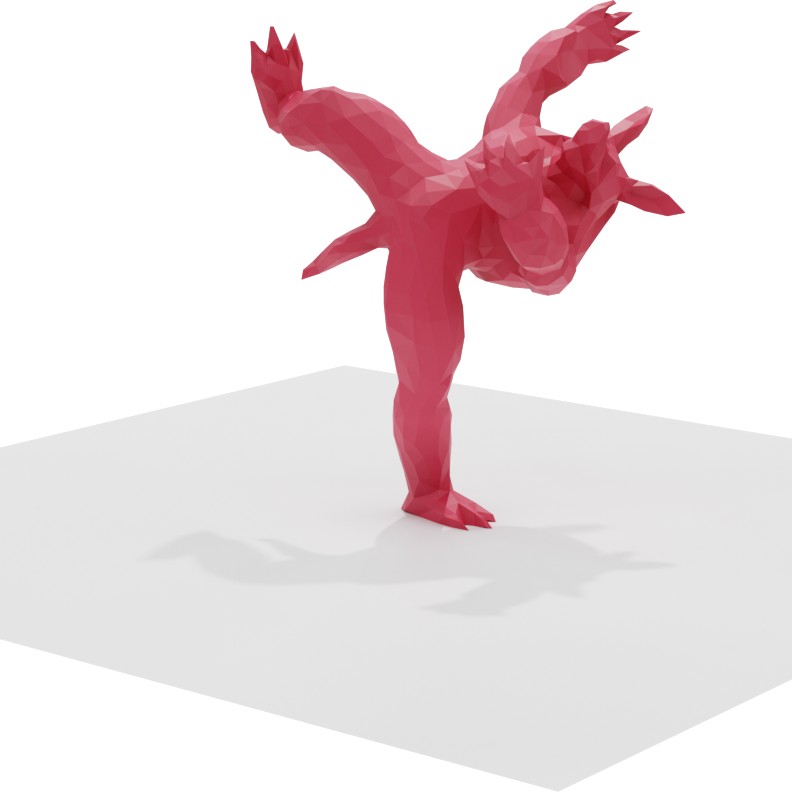} &
        \includegraphics[width=.22\linewidth]{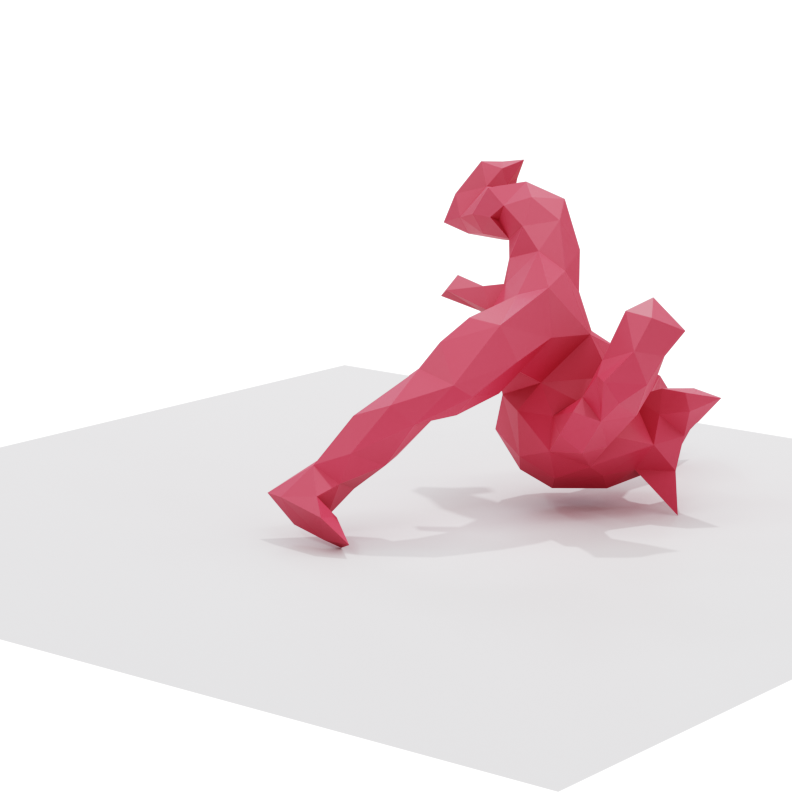} &
        \includegraphics[width=.22\linewidth]{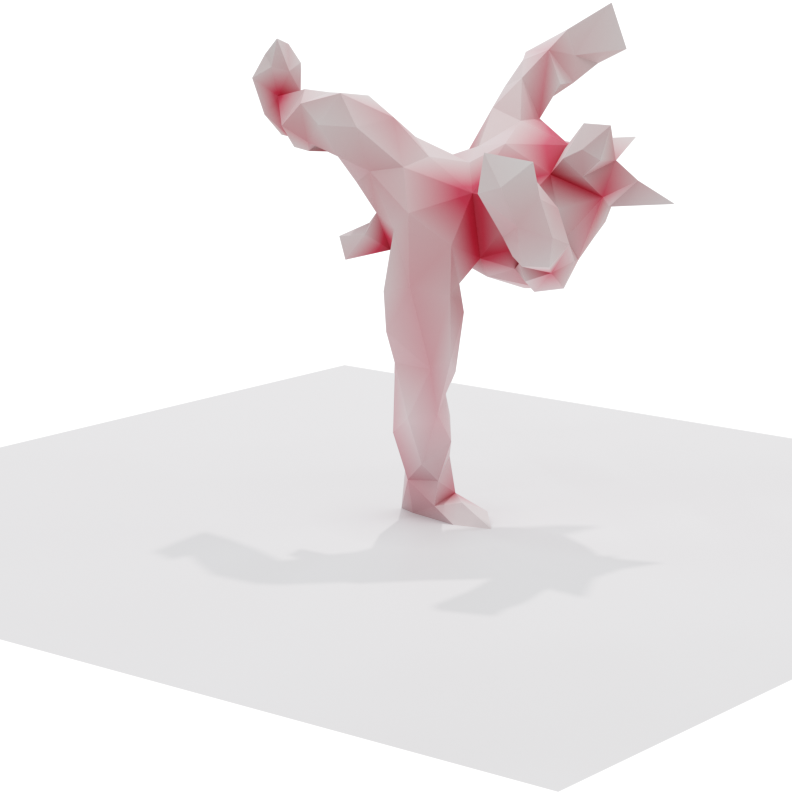}
    \end{tabular}
    \vspace{-2mm}
    \caption[Balancing armadillo]{\figname{Balancing armadillo.} We simulate the dancing armadillo from~\cite{Prevost2013Make} falling on a plane (left). The coarse model (middle) tips over because the center of mass falls outside the foot. We optimize the density (shown in red) to match the input center of mass and the armadillo is balanced (right). Differences in running time can be attributed to the different dynamics (i.e., the coarse model experiences more contacts when it falls over).}
    \label{fig:arma-balance}
    \vspace{-2mm}
\end{figure}

%% file: 60-concluding.tex
\section{Concluding Remarks}

We introduce a robust and efficient simulator for deformable objects with contact supporting high-order meshes and high-order bases to simulate geometrically complex scenes. We show that there are major computational advantages in increasing the order of the geometric map and bases and that they can be used in the \ac{IPC} formulation with modest code changes.

\paragraph{Limitations}
At a high level, we are proposing to use $p$-refinement for elasticity, coupled with $h$-refinement approach for contacts, to sidestep the high computational cost of curved continuous collision detection. The downside of our approach is that our contact surface is still an approximation of the curved geometry, and while we can reduce the error by further refinement, we cannot reduce it to zero. While for graphics applications this is an acceptable compromise, as the scene we use for collision is guaranteed to be collision-free and we inherit the robustness properties of the original \ac{IPC} formulation, there could be engineering applications where it is important to model a high-order surface \emph{exactly}. In this case, our approach could not be used as we might miss the collisions of the curved \ac{FE} mesh.

A second limitation of our approach is that the definition of a robust, guaranteed positivity check for high-order elements is still an open research problem \cite{Johnen2013Geometrical}. In our implementation, we check positivity only at the quadrature points, which is a reasonable approximation but might still lead to unphysical results as the element might have a negative determinant in other interior points.

While our method for mapping between an arbitrary triangle mesh proxy and the curved tetrahedral mesh works well enough for the examples shown in this paper, it is not a robust implementation, as the closest point query can lead to wrong correspondences. In the future, it will be interesting to explore the use of bijective maps between the two geometries to avoid this issue (for example by using the work of~\citet{Jiang2020Bijective}).

Our choice of $\Phi$ is not unique as there are a large number of basis functions to choose from. We explored other options such as mean value coordinates and linearized L2-projection, but we found their global mappings produce dense weight matrices. This results in slower running times with only minor quality improvements. A future direction might be the exploration of more localized operators such as bounded bi-harmonic weights~\cite{Jacobson2011Bounded}.

\paragraph{Future Work}
Beyond these limitations, we see three major avenues for future work: (1) existing curved mesh generators are still not as reliable in producing high-quality meshes as their linear counterparts: more work is needed in this direction, and our approach can be used as a testbed for evaluating the benefits curved mesh provides in the context of elastodynamic simulations, (2) our approach could be modified to work with hexahedral elements, spline bases, and isogeometric analysis simulation frameworks, and (3) we speculate that integrating our approach with high-order time integrators could provide additional benefits for further reducing numerical damping and we believe this is a promising direction for a future study.

Our approach is a first step toward the introduction of high-order meshes and high-order \ac{FEM} in elastodynamic simulation with the \ac{IPC} contact model, and we believe that our reference implementation will reduce the entry barrier for the use of these approaches in industry and academia.

%% file: 65-acknowledgments.tex
\begin{acks}
This work was supported in part through the NYU IT High Performance Computing resources, services, and staff expertise.
This work was also partially supported by the NSF CAREER award under Grant No. 1652515, the NSF grants OAC-1835712, OIA-1937043, CHS-1908767, CHS-1901091, NSERC DGECR-2021-00461 and RGPIN 2021-03707, a Sloan Fellowship, a gift from Adobe Research and a gift from Advanced Micro Devices, Inc.
\end{acks}

%% file: 70-figures.tex
\input{figs/rolling-ball-friction.tex}
\input{figs/bolt.tex}
\input{figs/trash-compactor.tex}
\input{figs/microstructure.tex}

\input{figs/armadillo-rollers.tex}

\input{figs/sim_params.tex}
\input{figs/sim_results.tex}

\clearpage

%% file: figs/rolling-ball-friction.tex
\begin{figure}[!b]
\centering\footnotesize
\includegraphics[width=\linewidth]{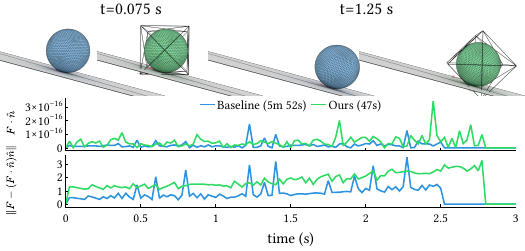}
\vspace{-6mm}
\caption[Rolling-ball]{\figname{Rolling-ball.} We demonstrate a ball rolling down a slope, while maintaining non-slip rolling contact, produces purely tangential friction forces on the FEM mesh. Our method uses a symmetric cube mesh (black wireframe) as the \ac{FE} mesh and a high-resolution sphere (green) as the collision mesh. The friction forces on the \ac{FE} mesh are shown as pink arrows. We plot the out-of-plane friction force ($F \cdot \hat{n}$) and norm of the in-plane friction force ($\|F - (F\cdot \hat{n})\hat{n}\|$). Compared to a high-resolution baseline, the out-of-plane error shows negligible differences but the in-plane force is around $2\times$ greater. This is due to the increased numerical stiffness of our course mesh leading to less localized deformation, smaller distances, and, ultimately, a larger normal force.}
\label{fig:rolling_ball_friction}
\end{figure}

%% file: figs/bolt.tex
\begin{figure}
    \centering\footnotesize
    \parbox{.06\linewidth}{~}\hfill
    \parbox{.3\linewidth}{\raggedleft $t=\qty{0.0}{\s}$\hspace{0.75em}\ }\hfill
    \parbox{.3\linewidth}{\raggedleft $t=\qty{2.0}{\s}$\hspace{0.75em}\ }\hfill
    \parbox{.3\linewidth}{\raggedleft $t=\qty{5.0}{\s}$\hspace{0.75em}\ }\\

    \parbox{.06\linewidth}{\centering\rotatebox[origin=c]{90}{\parbox{2.5\linewidth}{\centering Baseline\\ (\ms{22}{4})}}}\hfill
    \parbox{.3\linewidth}{\includegraphics[width=\linewidth]{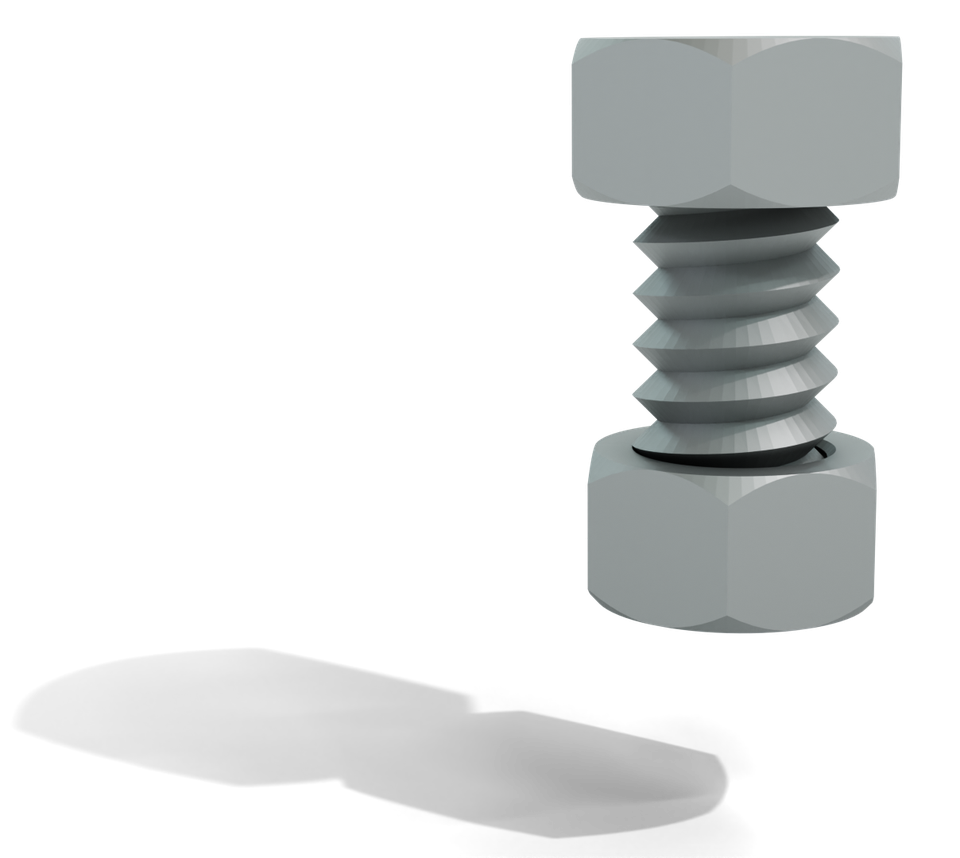}}\hfill
    \parbox{.3\linewidth}{\includegraphics[width=\linewidth]{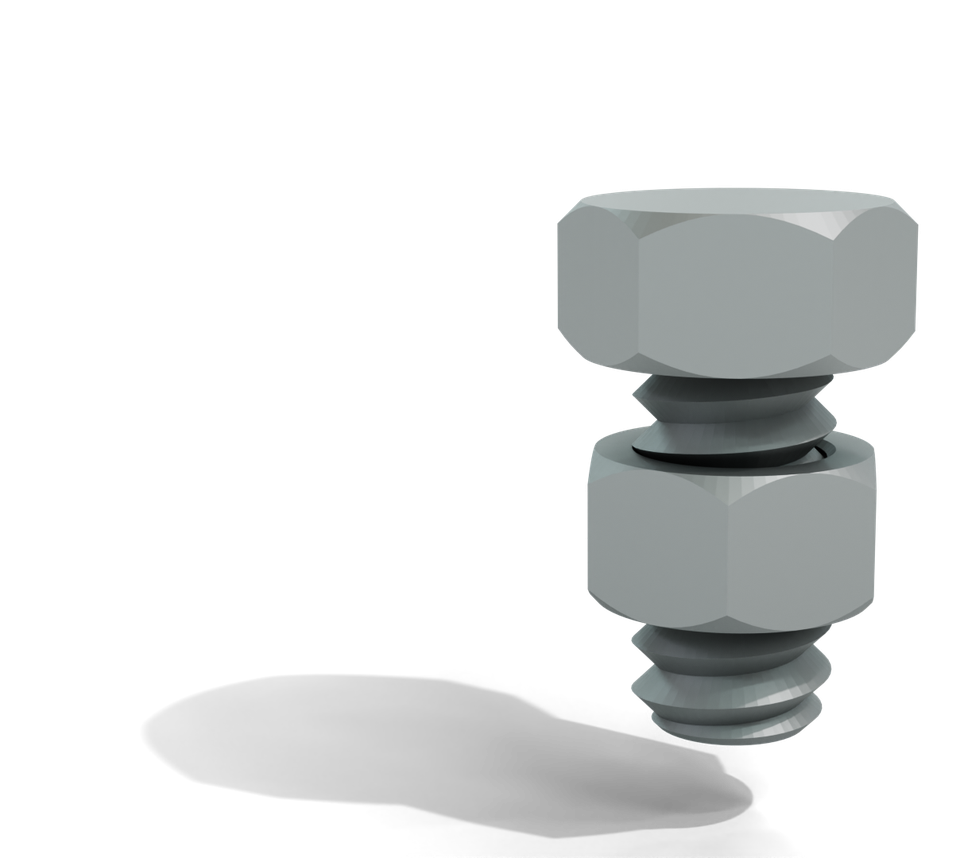}}\hfill
    \parbox{.3\linewidth}{\includegraphics[width=\linewidth]{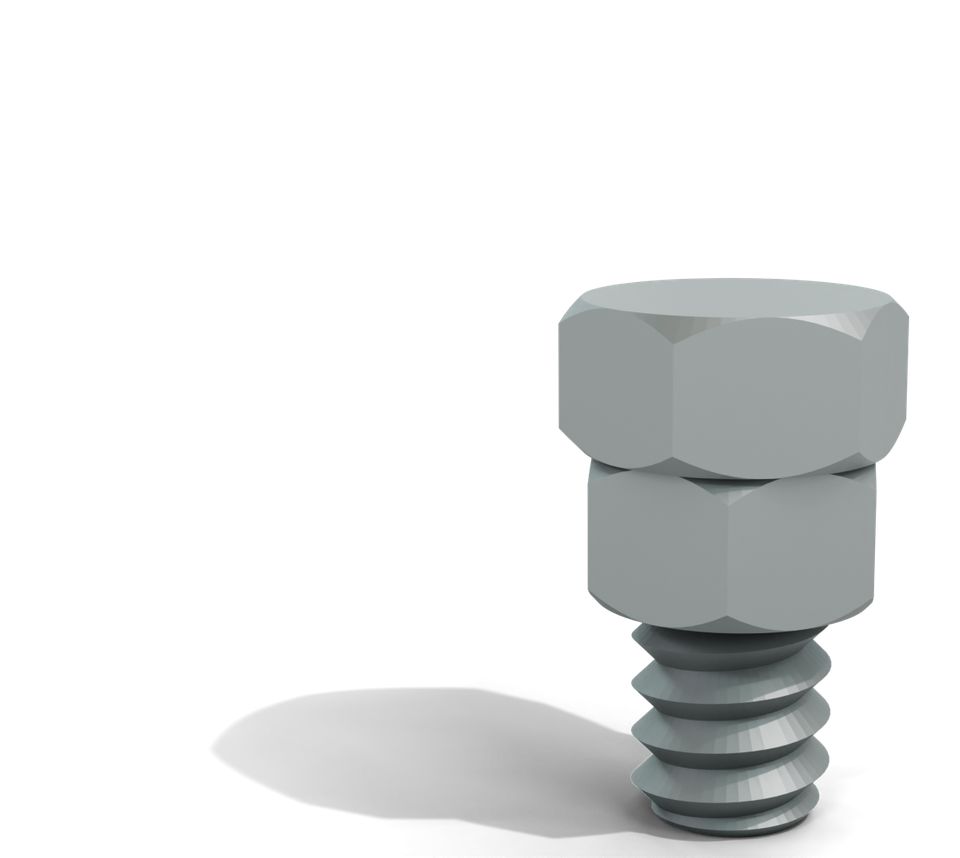}}\\

    \parbox{.06\linewidth}{\centering\rotatebox[origin=c]{90}{\parbox{2.5\linewidth}{\centering Ours\\ (\ms{9}{40})}}}\hfill
    \parbox{.3\linewidth}{\includegraphics[width=\linewidth]{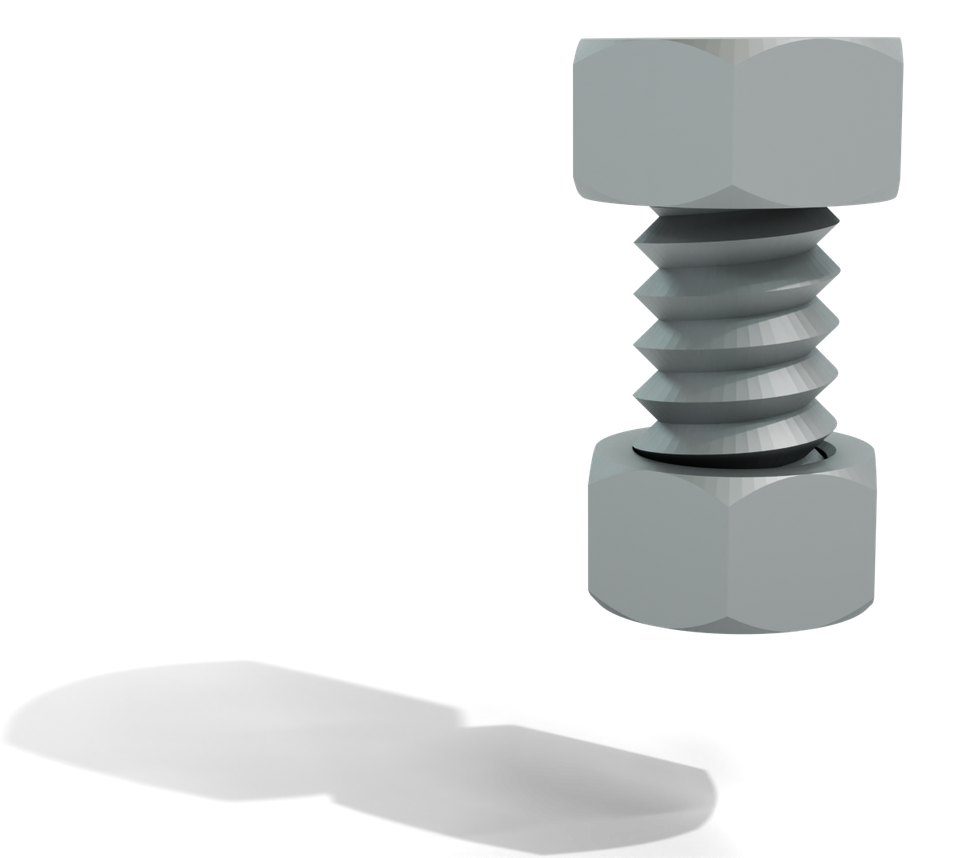}}\hfill
    \parbox{.3\linewidth}{\includegraphics[width=\linewidth]{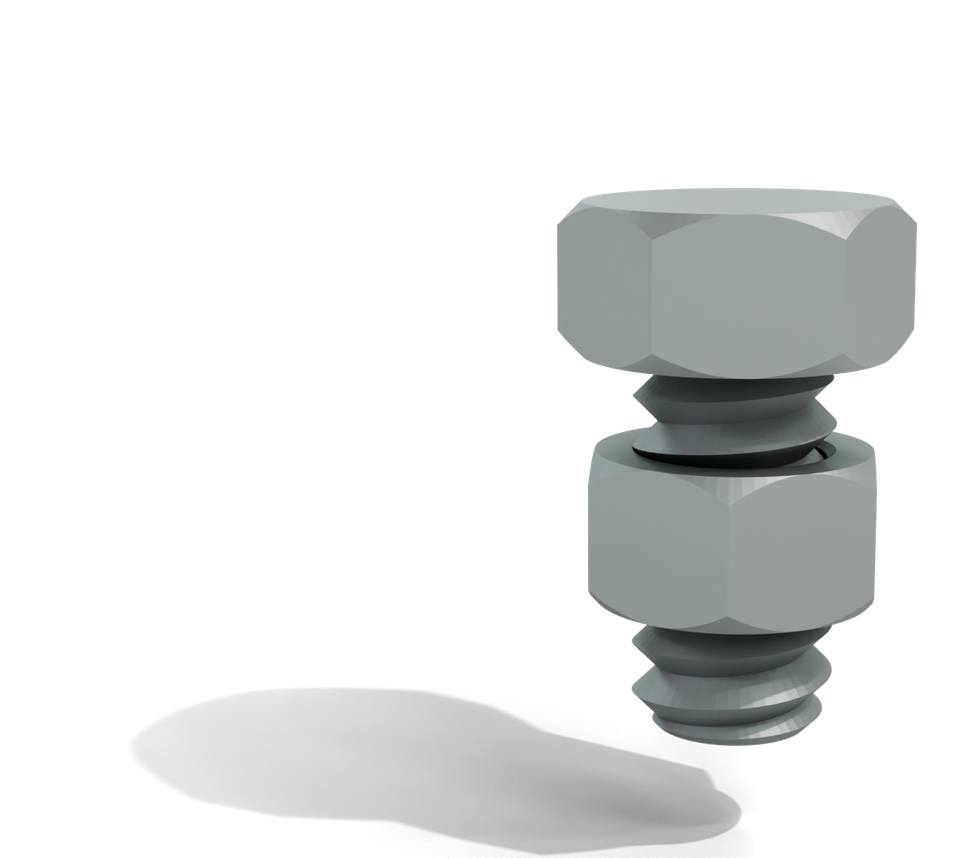}}\hfill
    \parbox{.3\linewidth}{\includegraphics[width=\linewidth]{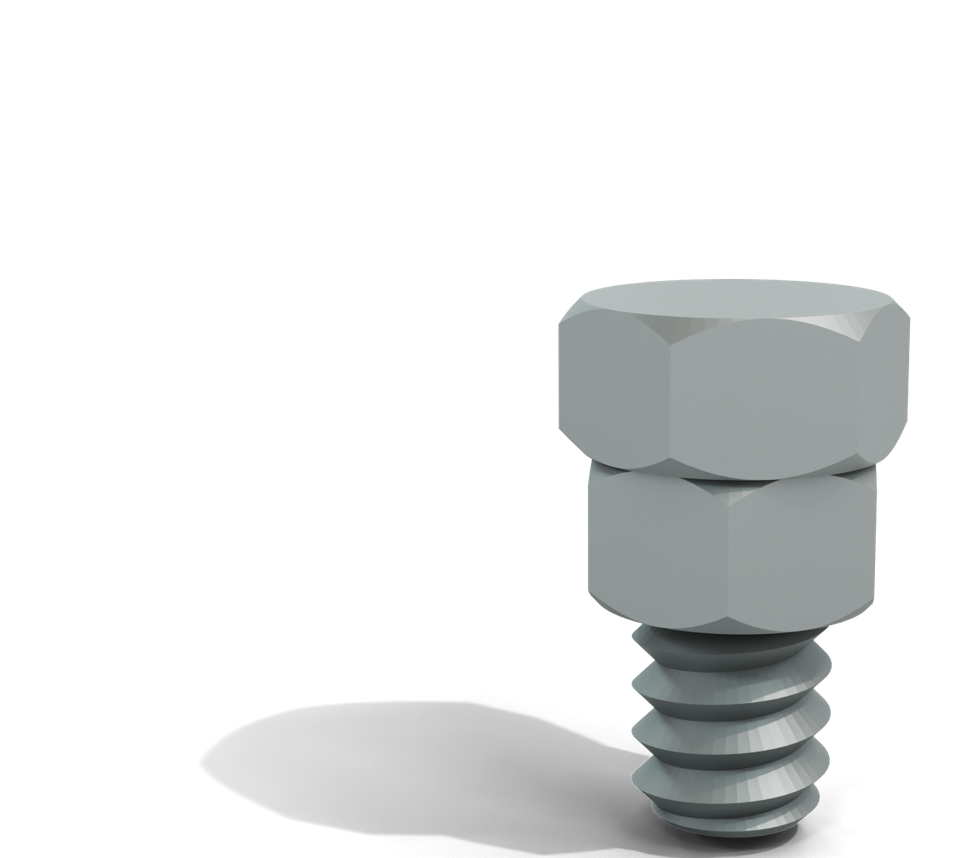}}\\

    \parbox{.06\linewidth}{\centering\rotatebox[origin=c]{90}{Coarse FE Mesh}}
    \parbox{0.3\linewidth}{\includegraphics[width=\linewidth]{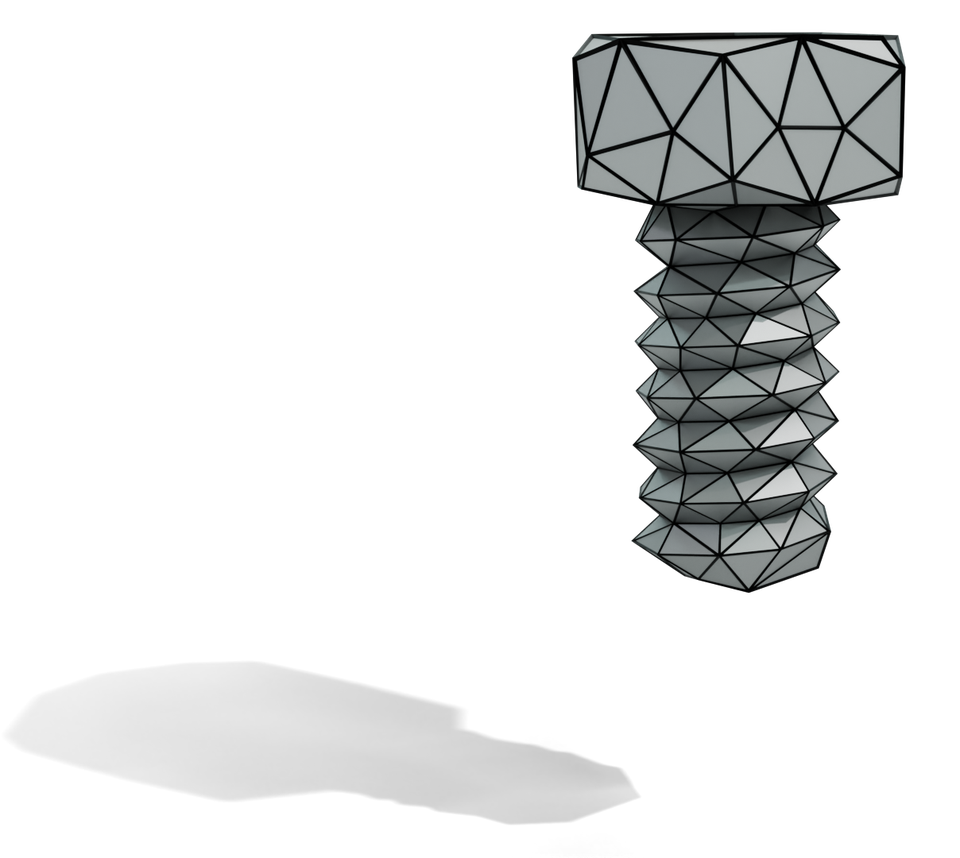}}\\
    \copyrightmodel{YSoft be3D}{CC BY-SA $3.0$}
    \vspace{-4mm}
    \caption[Nut-and-bolt]{\figname{Nut-and-bolt.} Simulation of a bolt rotating into a bolt under gravity. Directly meshing the input mesh (top) generate similar results as using our method with a coarse simulation mesh (right).}
    \label{fig:screw}
\end{figure}

%% file: figs/trash-compactor.tex
\begin{figure}
    \centering\footnotesize
    \parbox{\linewidth}{\centering
    \parbox{.06\linewidth}{~}\hfill
    \parbox{.3\linewidth}{\centering$t=\qty{0.5}{\s}$}\hfill
     \parbox{.3\linewidth}{\centering$t=\qty{0.8}{\s}$}\hfill
     \parbox{.3\linewidth}{\centering$t=\qty{1.8}{\s}$}\\

    \parbox{.06\linewidth}{\centering\rotatebox[origin=c]{90}{\parbox{2.5\linewidth}{\centering Baseline\\ (\hms{5}{8}{25})}}}\hfill
    \parbox{.3\linewidth}{\includegraphics[width=\linewidth]{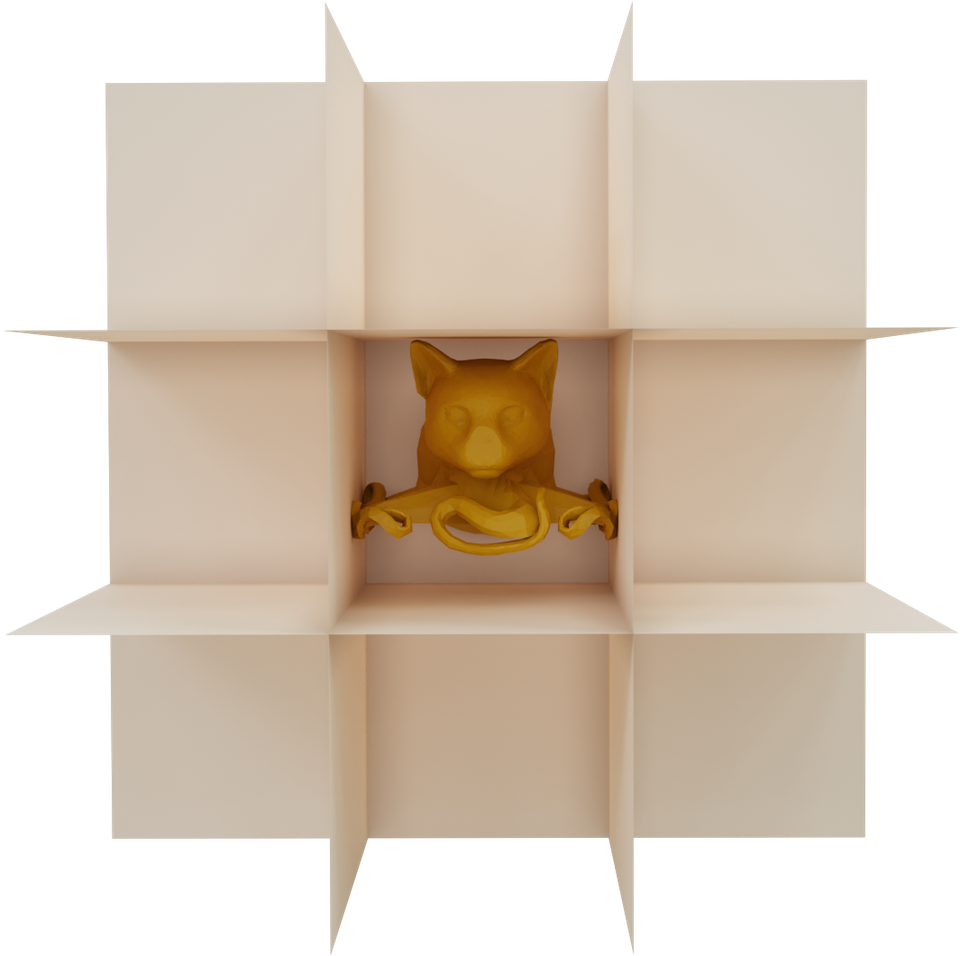}}\hfill
    \parbox{.3\linewidth}{\includegraphics[width=\linewidth]{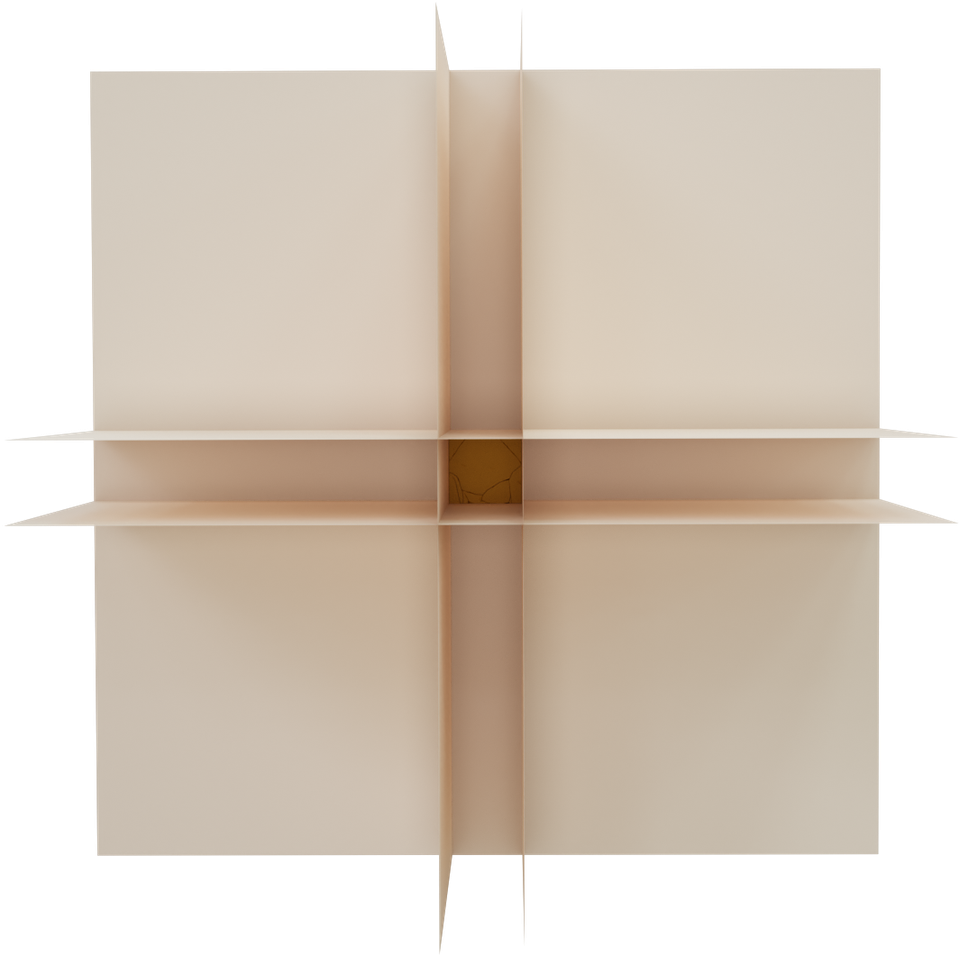}}\hfill
    \parbox{.3\linewidth}{\includegraphics[width=\linewidth]{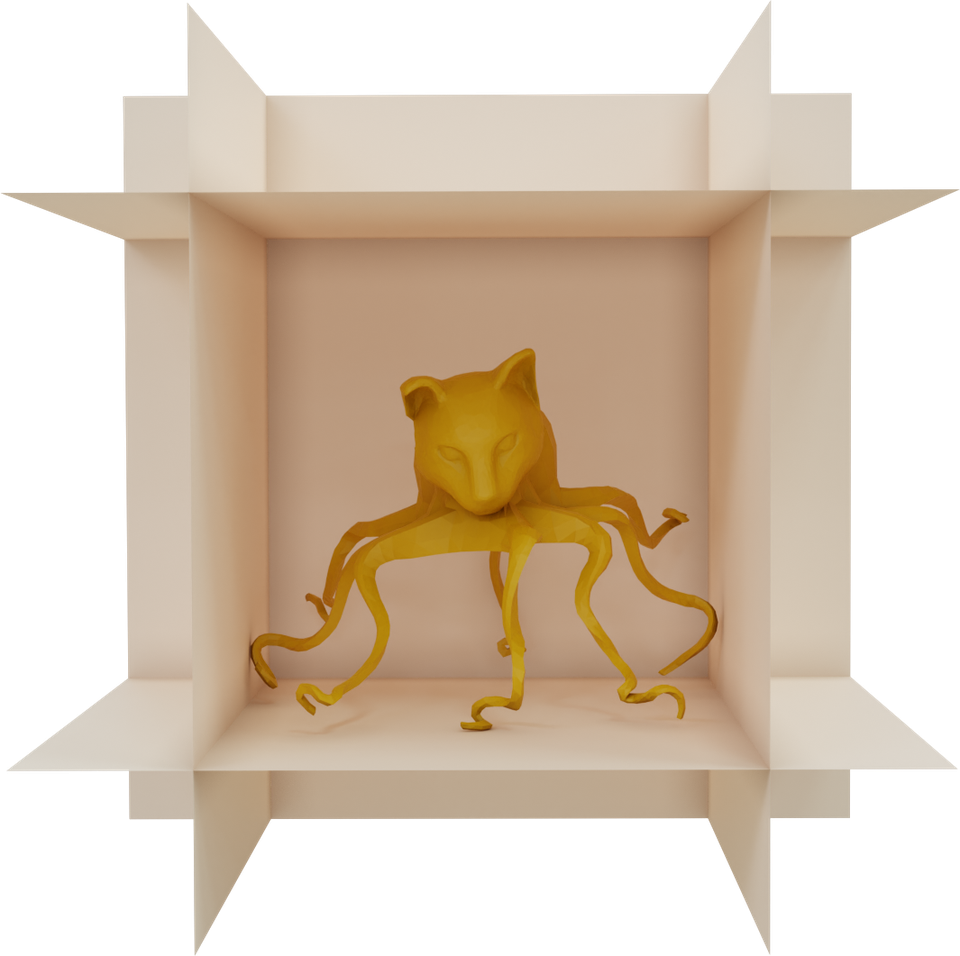}}\\

    \parbox{.06\linewidth}{\centering\rotatebox[origin=c]{90}{\parbox{2.5\linewidth}{\centering Ours\\ (\hms{2}{20}{16})}}}\hfill
    \parbox{.3\linewidth}{\includegraphics[width=\linewidth]{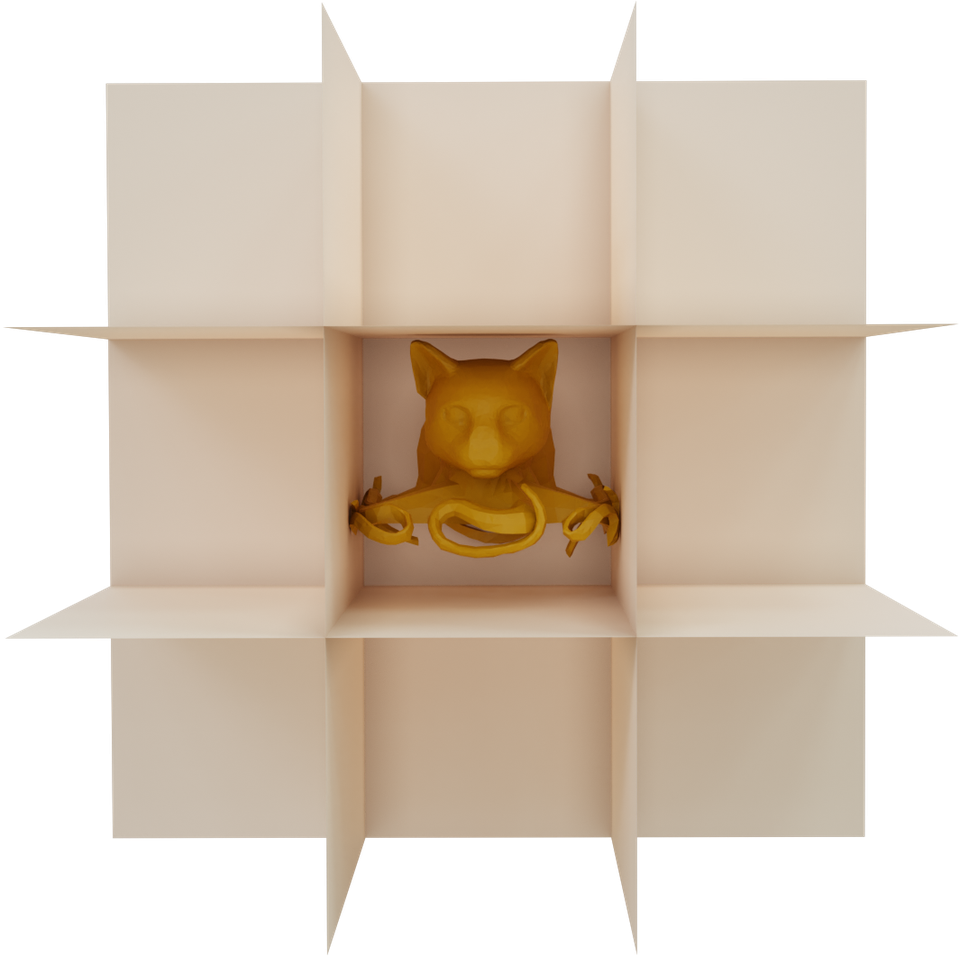}}\hfill
    \parbox{.3\linewidth}{\includegraphics[width=\linewidth]{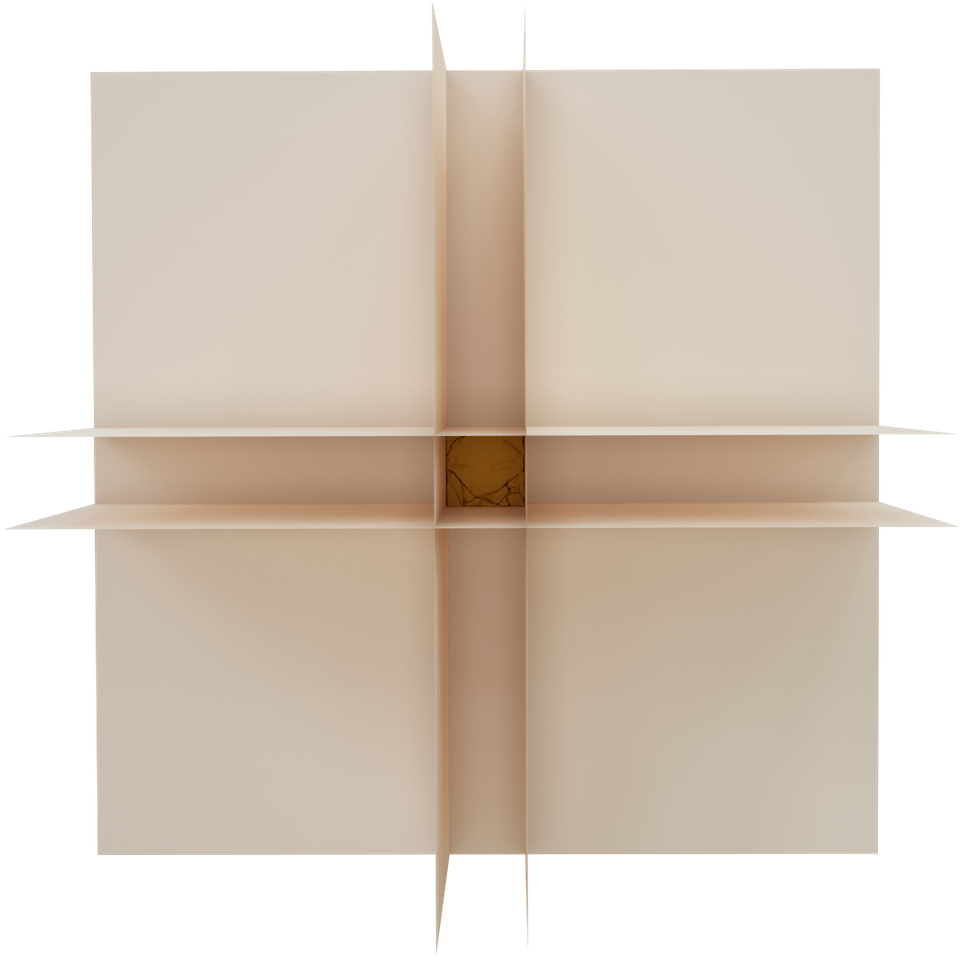}}\hfill
    \parbox{.3\linewidth}{\includegraphics[width=\linewidth]{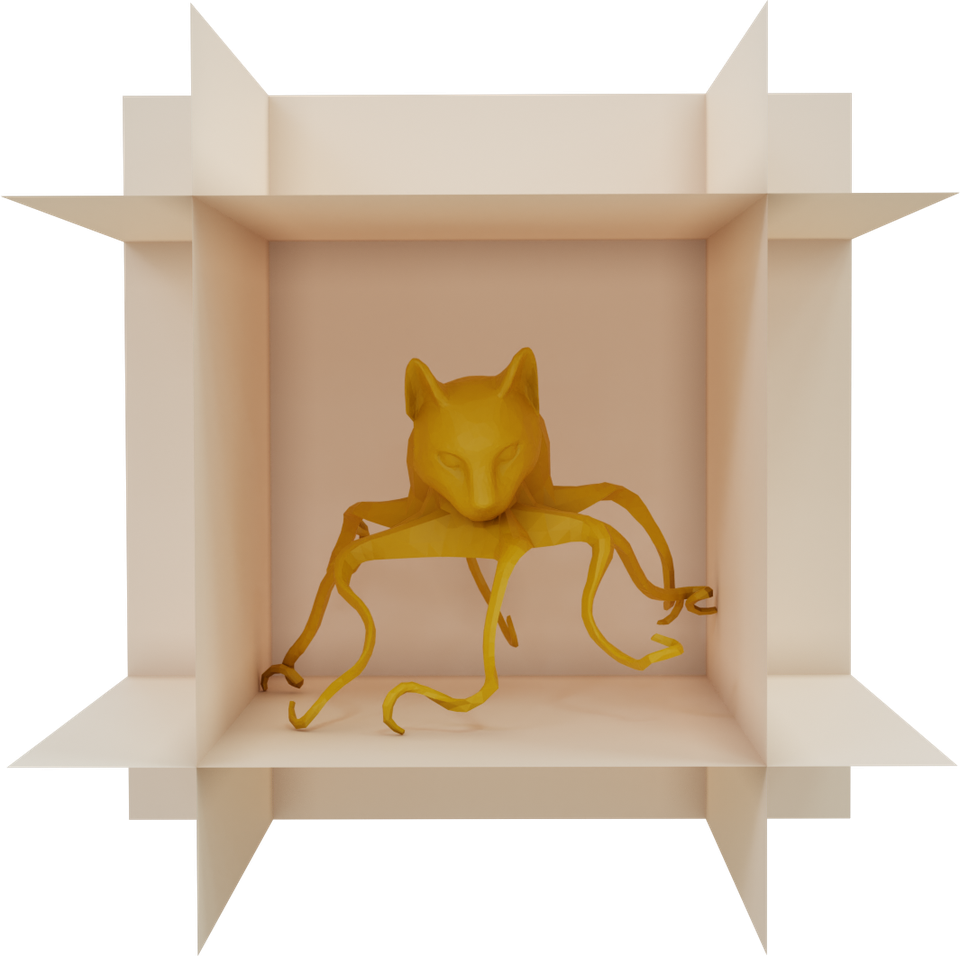}}\\

    \parbox{.06\linewidth}{\centering\rotatebox[origin=c]{90}{\parbox{2.5\linewidth}{\centering Ours, $P_2$\\ (\hms{24}{23}{0})}}}\hfill
    \parbox{.3\linewidth}{\includegraphics[width=\linewidth]{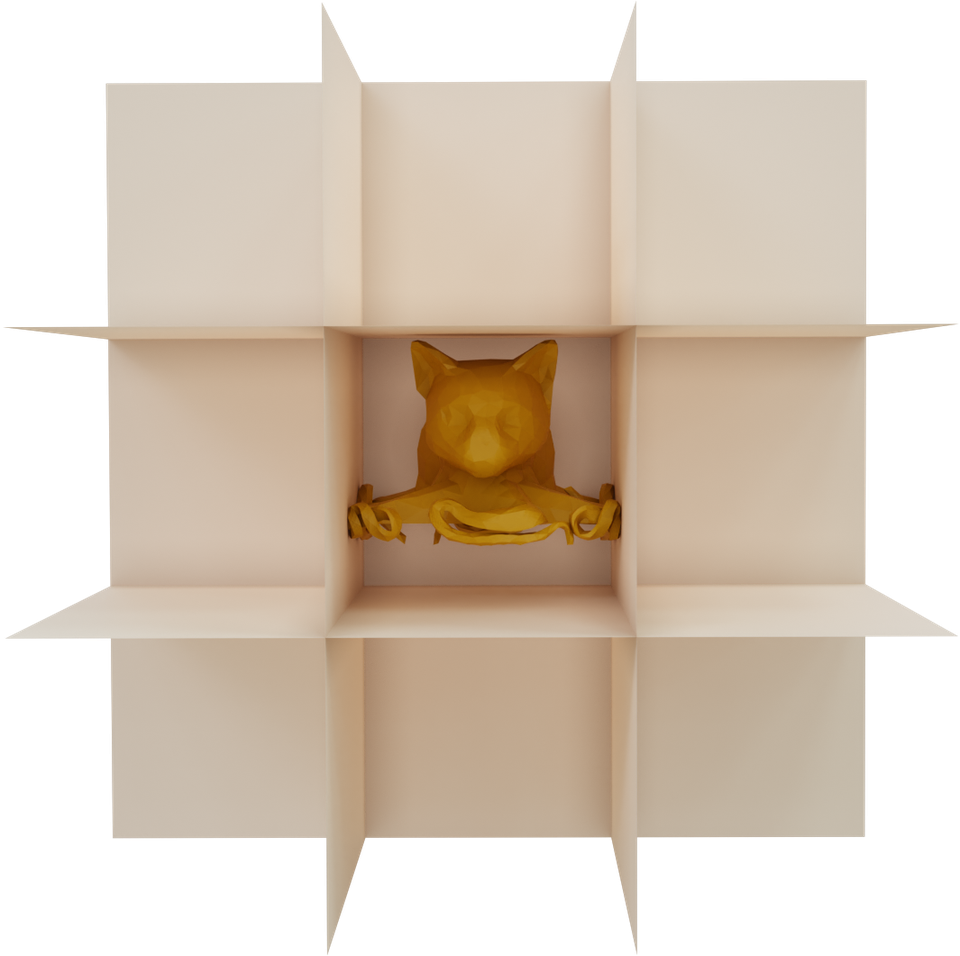}}\hfill
    \parbox{.3\linewidth}{\includegraphics[width=\linewidth]{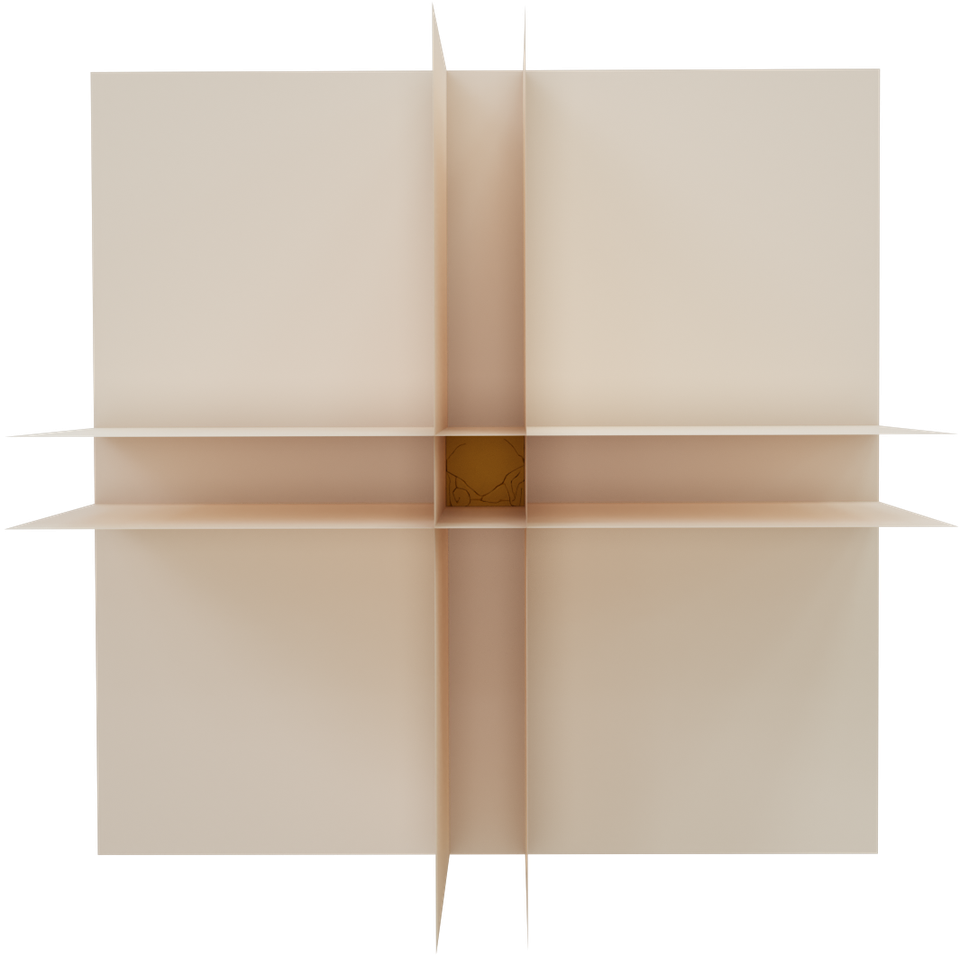}}\hfill
    \parbox{.3\linewidth}{\includegraphics[width=\linewidth]{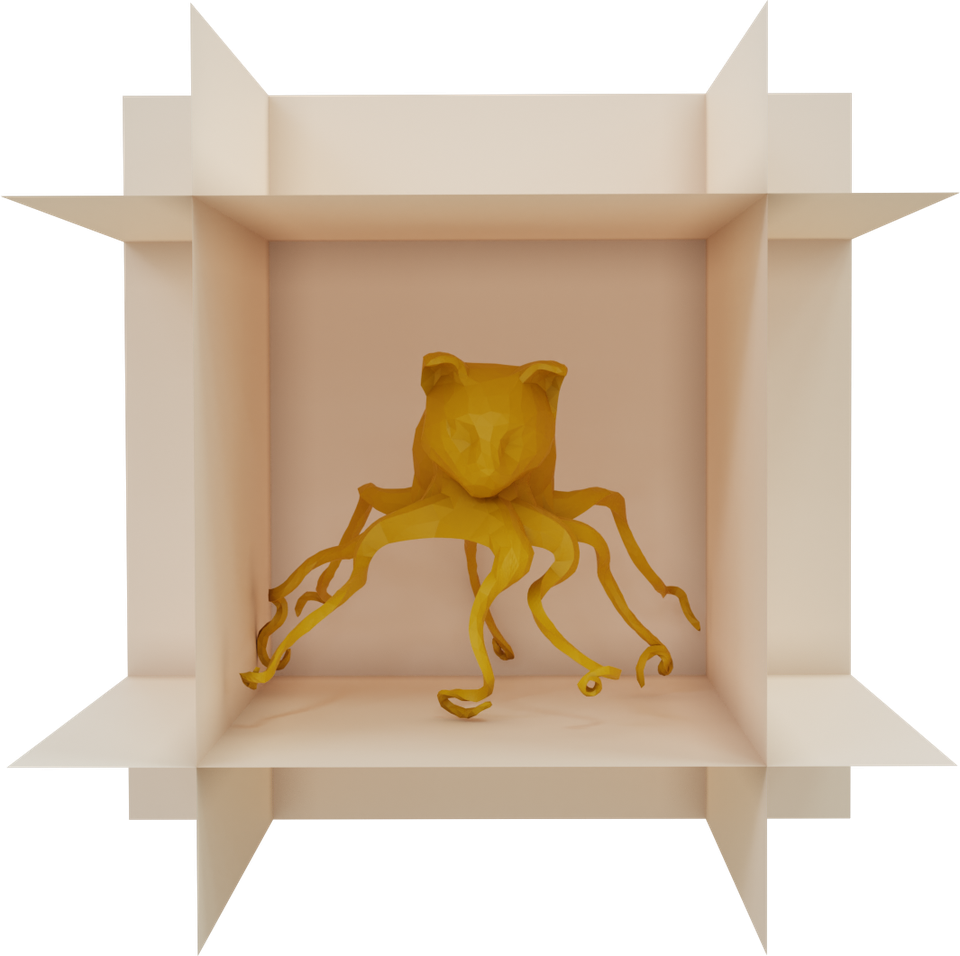}}\\

    \parbox{.06\linewidth}{\centering\rotatebox[origin=c]{90}{Coarse FE Mesh}}
    \parbox{.3\linewidth}{\includegraphics[width=\linewidth]{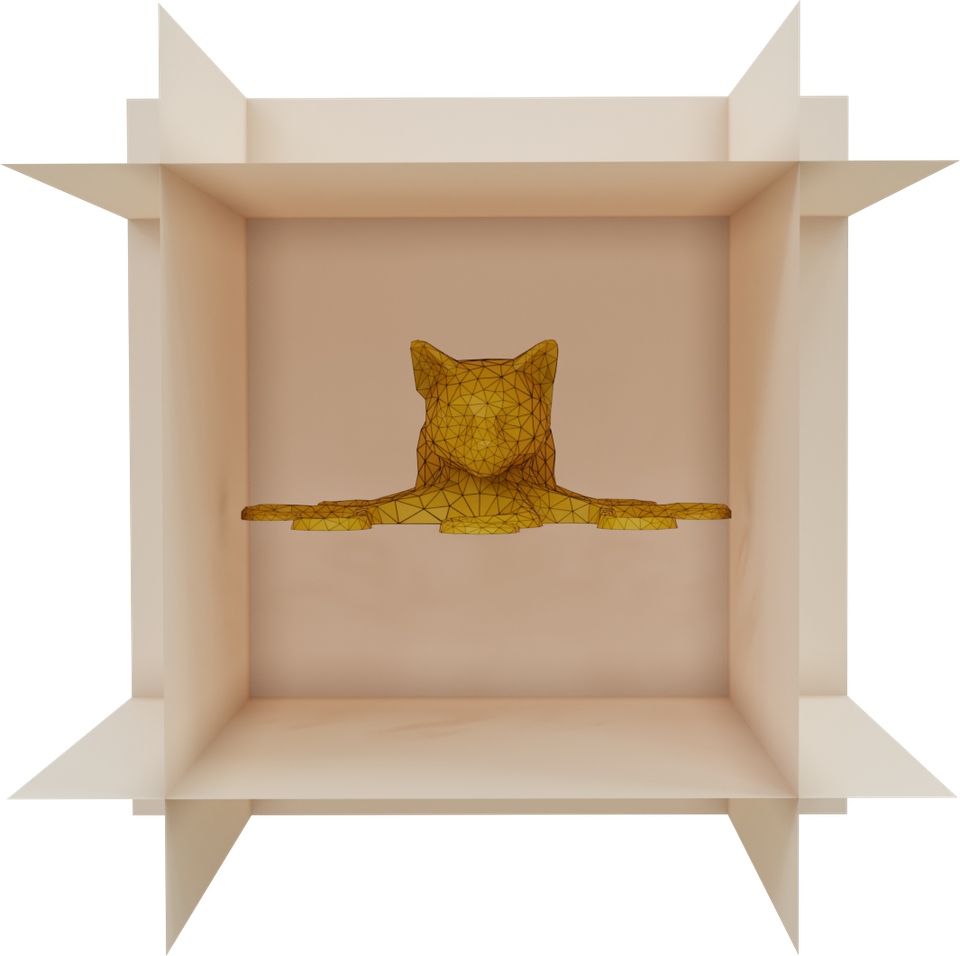}}}

    \copyrightmodel{Brian Enigma}{CC BY-SA 3.0}
    \vspace{-4mm}
    \caption[Trash compactor]{\figname{Trash compactor.} The Octocat model is compressed by five planes. Using the original input mesh (top) is two times slower than using our method with linear elements (middle). Since we cannot coarsen the input too much without losing the tentacles, using $P_2$ leads to longer running times and similar results (bottom). }
    \label{fig:trash-compactor}
\end{figure}

%% file: figs/microstructure.tex
\begin{figure}
    \centering\footnotesize
    \parbox{.06\linewidth}{~}\hfill
    \parbox{.3\linewidth}{\centering$t=\qty{0.2}{\s}$}\hfill
     \parbox{.3\linewidth}{\centering$t=\qty{0.4}{\s}$}\hfill
     \parbox{.3\linewidth}{\centering$t=\qty{1.0}{\s}$}\\

    \parbox{.06\linewidth}{\centering\rotatebox[origin=c]{90}{\parbox{2.5\linewidth}{\centering $P_1$\\ (\hms{6}{34}{9})}}}\hfill
    \parbox{.3\linewidth}{\includegraphics[width=\linewidth]{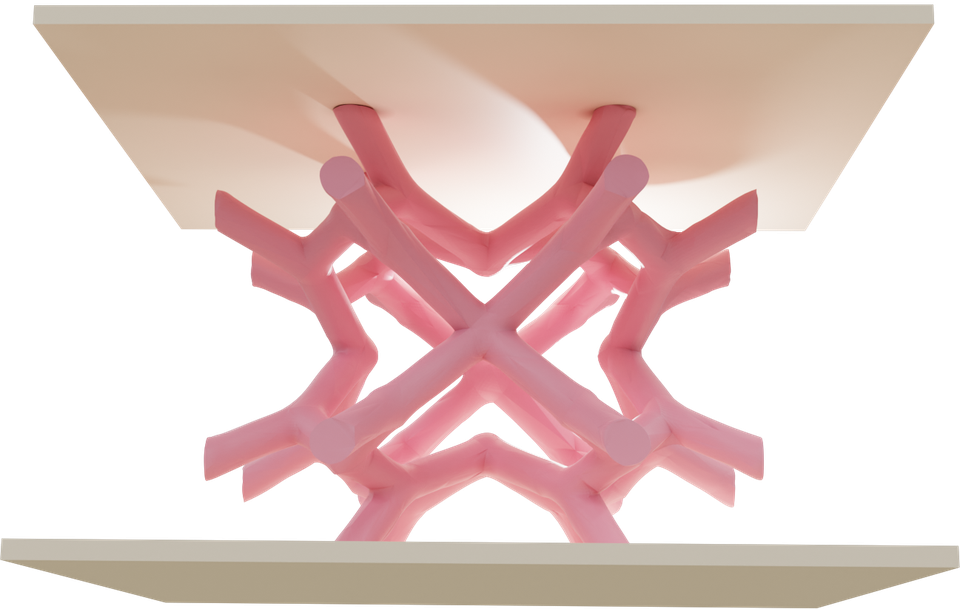}}\hfill
    \parbox{.3\linewidth}{\includegraphics[width=\linewidth]{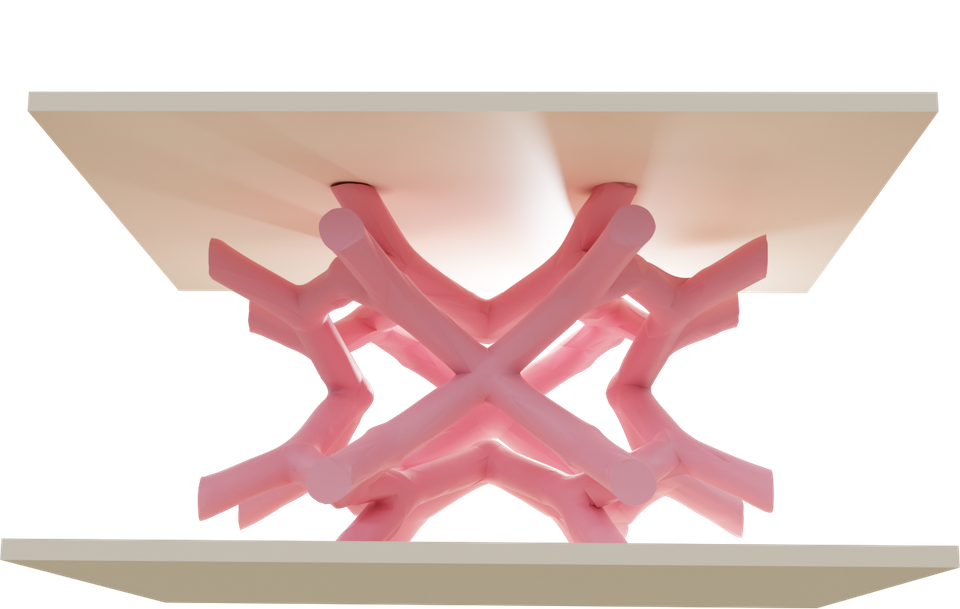}}\hfill
    \parbox{.3\linewidth}{\includegraphics[width=\linewidth]{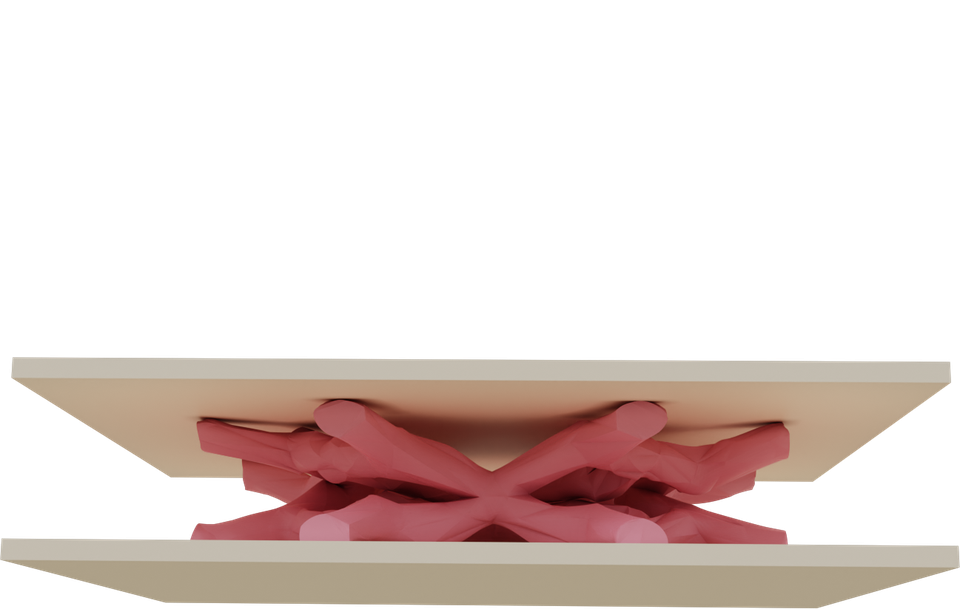}}\\

    \parbox{.06\linewidth}{\centering\rotatebox[origin=c]{90}{\parbox{2.5\linewidth}{\centering $P_2$\\ (\hms{6}{4}{48})}}}\hfill
    \parbox{.3\linewidth}{\includegraphics[width=\linewidth]{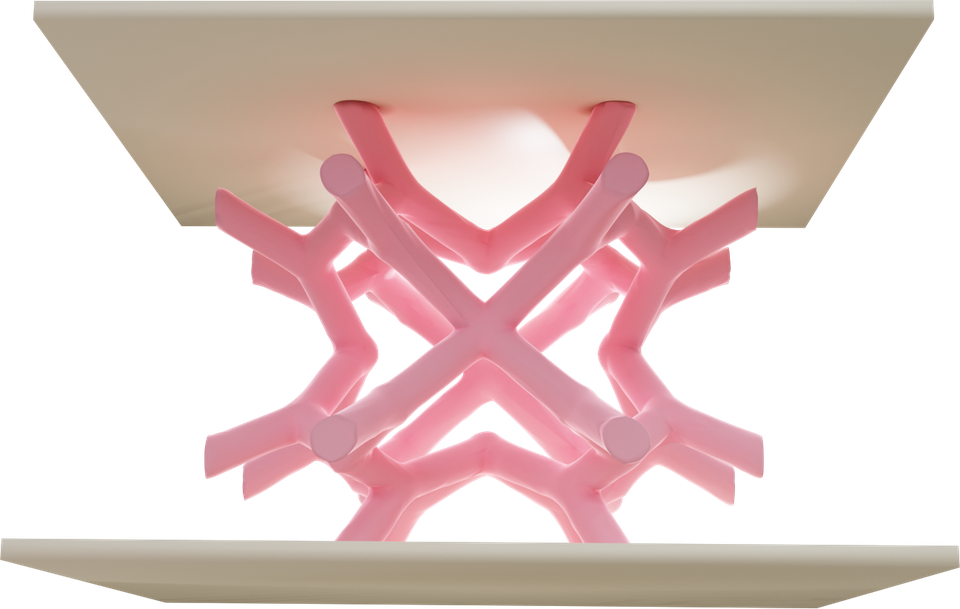}}\hfill
    \parbox{.3\linewidth}{\includegraphics[width=\linewidth]{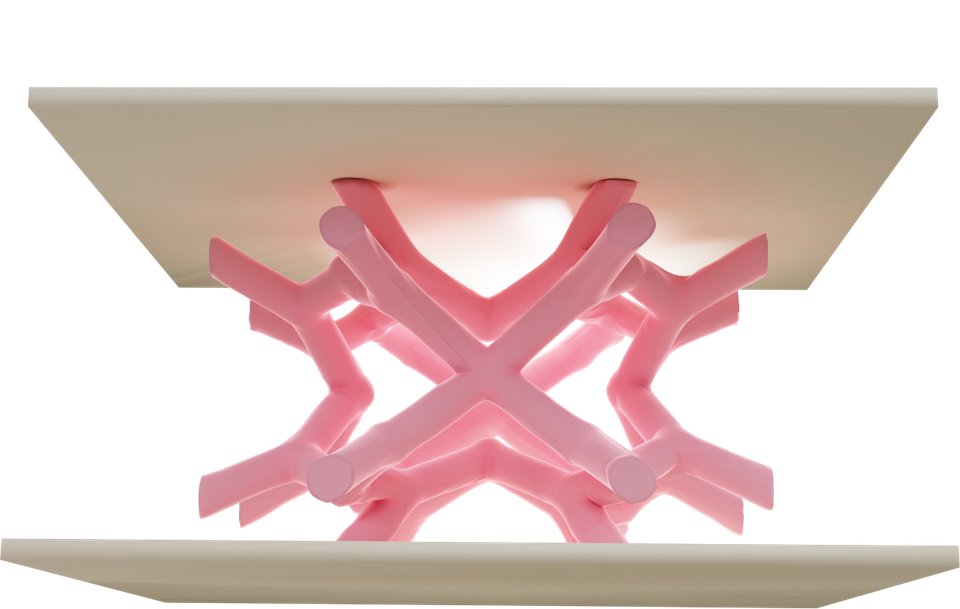}}\hfill
    \parbox{.3\linewidth}{\includegraphics[width=\linewidth]{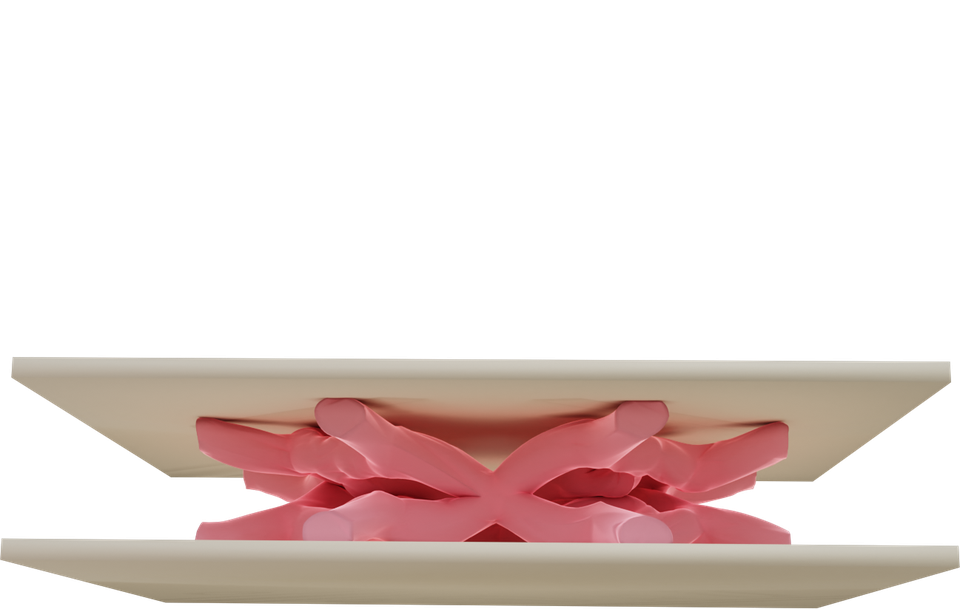}}\\
    \caption[Microstructure]{\figname{Microstructure.} Compression of a curved microstructure using linear and quadratic bases. While the choice of bases only leads to marginal running time savings, it demonstrates our method's ability to simulate anisoparametric scenarios where the $P_4$ shape functions differ from the $P_1$/$P_2$ bases.}
    \label{fig:microstructure}
\end{figure}

%% file: figs/armadillo-rollers.tex
\begin{figure}
    \centering\footnotesize
    \parbox{.06\linewidth}{~}
    \parbox{.3\linewidth}{\centering $t=\qty{1.0}{\s}$}
    \parbox{.3\linewidth}{\centering $t=\qty{4.0}{\s}$}
    \parbox{.3\linewidth}{\centering $t=\qty{6.25}{\s}$}\\
    \parbox{.06\linewidth}{\centering\rotatebox[origin=c]{90}{\parbox{2.5\linewidth}{\centering Baseline\\(\dhms{2}{13}{19}{0})}}}
    \parbox{0.9\linewidth}{\includegraphics[width=\linewidth]{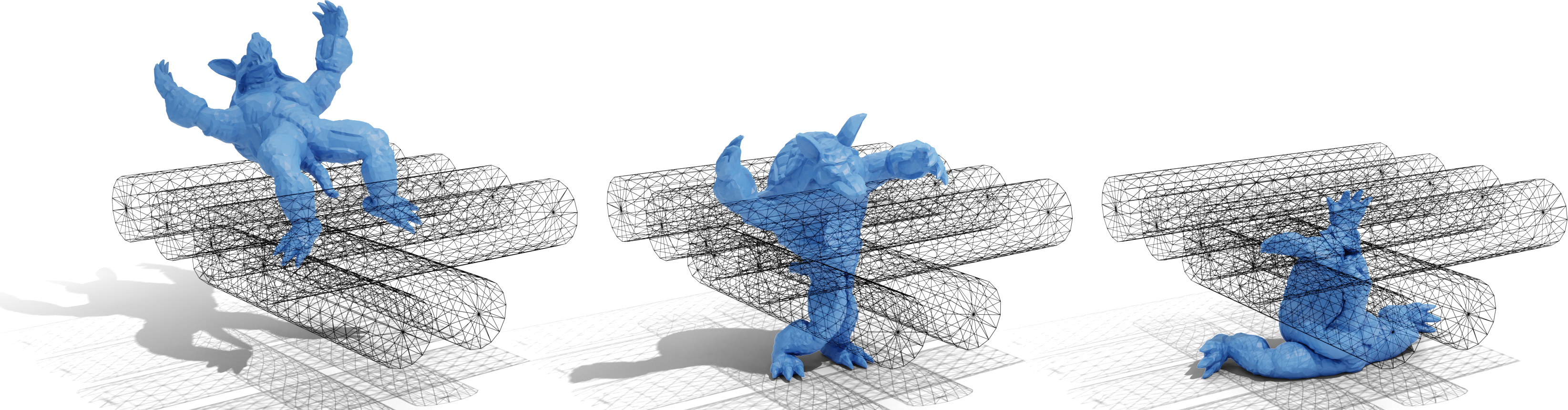}}\\[1mm]
    \parbox{.06\linewidth}{\centering\rotatebox[origin=c]{90}{\parbox{2.5\linewidth}{\centering Ours$^\star$, $P_1$\\(\ms{57}{36})}}}
    \parbox{0.9\linewidth}{\includegraphics[width=\linewidth]{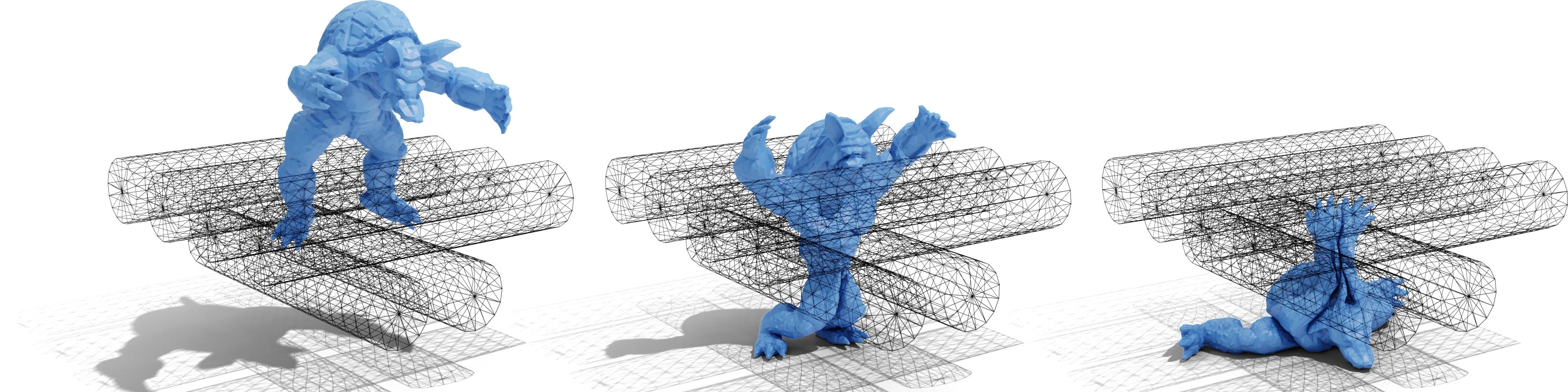}}\\[1mm]
    \parbox{.06\linewidth}{\centering\rotatebox[origin=c]{90}{\parbox{2.5\linewidth}{\centering Ours$^\dagger$, $P_2$\\(\hms{3}{58}{0})}}}
    \parbox{0.9\linewidth}{\includegraphics[width=\linewidth]{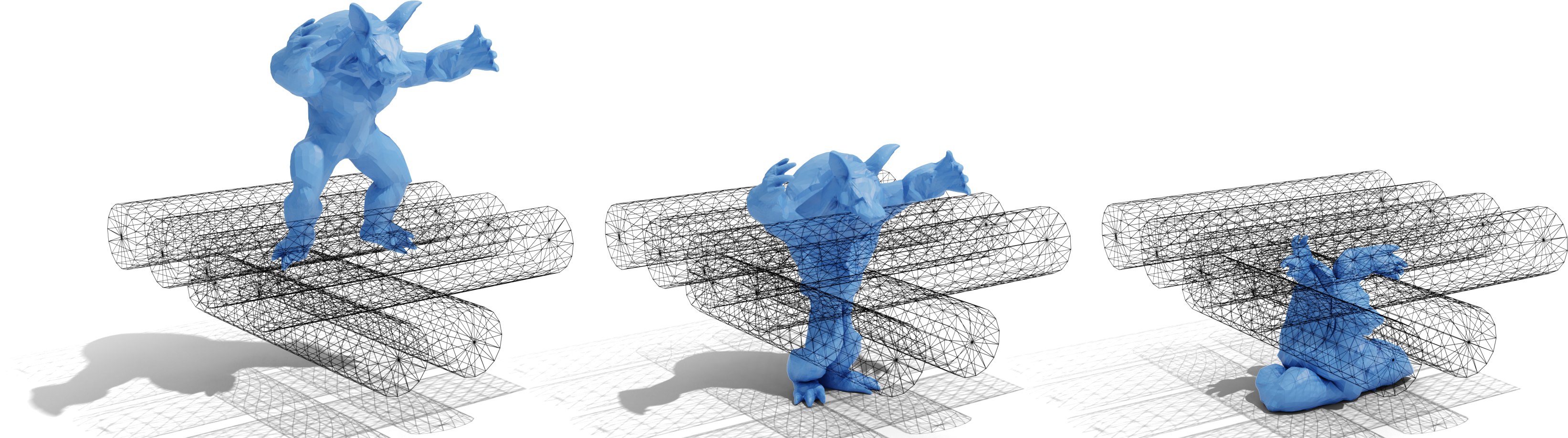}}\\[1mm]
    \parbox{.06\linewidth}{\centering\rotatebox[origin=c]{90}{\parbox{2.5\linewidth}{\centering Ours$^\ddagger$, $P_2$\\(\hms{7}{14}{32})}}}
    \parbox{0.9\linewidth}{\includegraphics[width=\linewidth]{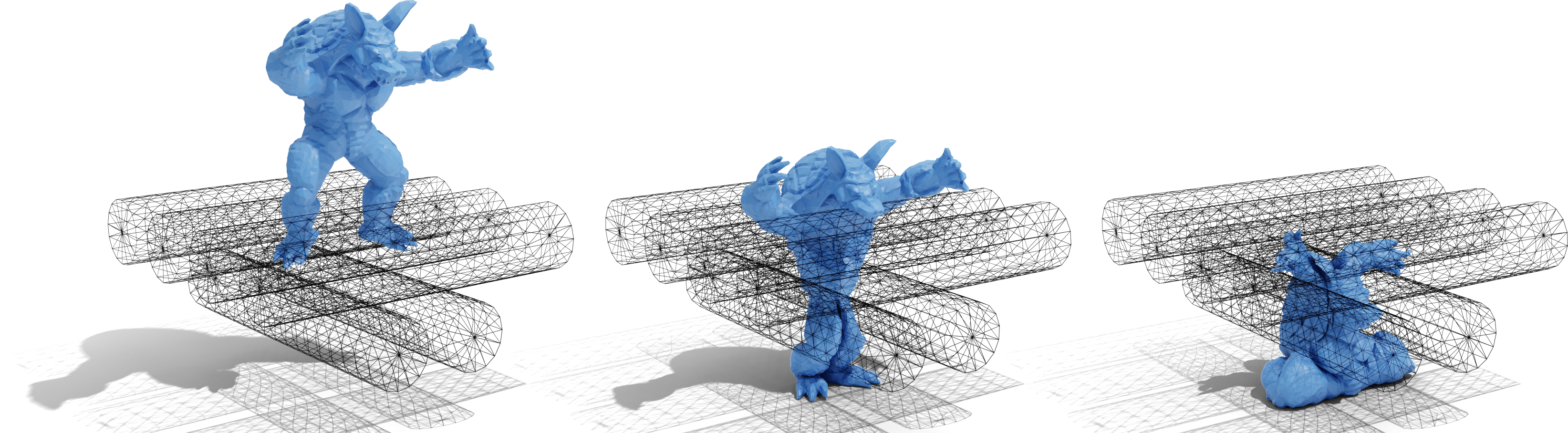}}\\[1mm]
    \parbox{.06\linewidth}{\centering\rotatebox[origin=c]{90}{Coarse FE Mesh}}
    \parbox{.45\linewidth}{\includegraphics[width=\linewidth]{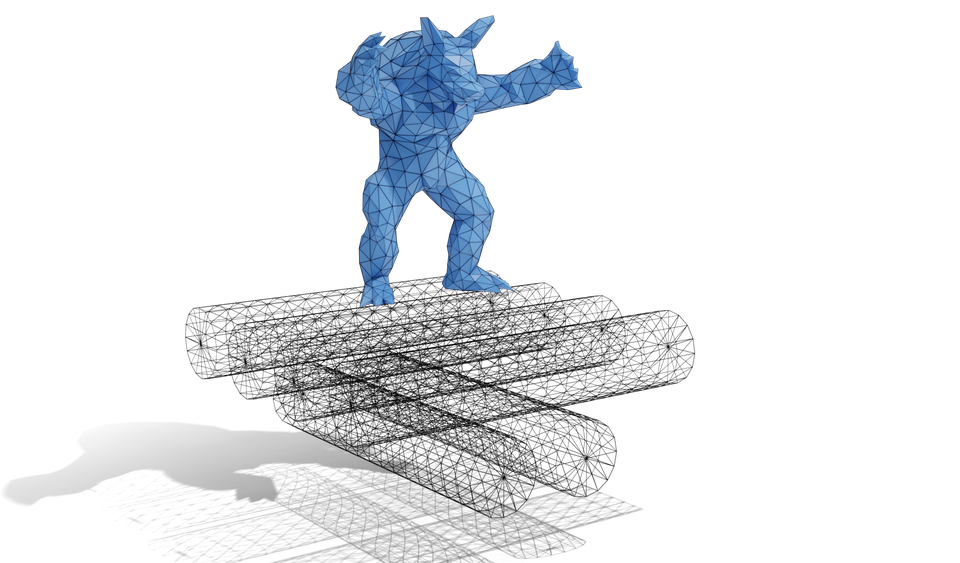}}
    \vspace{-2mm}
    \caption[Armadillo-rollers]{\figname{Armadillo-rollers.} Armadillo roller simulation for the different variants of our method. Ours$^\star$ uses a coarse linear mesh with linear displacement and the original geometry for the collision. Ours$^\dagger$ uses a curved mesh with $P_2$ displacement and an upsampled geometry for the collision. Ours$^\ddagger$ uses a curved mesh with $P_2$ displacement and the original geometry for the collision.}
    \label{fig:arma-roller-proxy}
\end{figure}

%% file: figs/sim_params.tex
\begin{table}
    \caption[High-order IPC simulation parameters]{Simulation parameters used in the results. For each example, we report the time step size ($h$), density ($\rho$ with $^\star$ indicating multi-density), Young's modulus ($E$), Poisson ratio ($\nu$), barrier activation distance ($\hat{d}$), coefficient of friction ($\mu$), friction smoothing parameter ($\epsilon_v$), maximum friction iterations, and Newton tolerance. For all examples, we use implicit Euler time integration and the Neo-Hookean material model.}
    \label{tab:sim_params}
    \centering
    \footnotesize
    \setlength{\tabcolsep}{4pt}
    \begin{tabular}{c|c|c|c|c|c}
        Scene & $h$ (\unit{\s}) & \makecell{$\rho$ (\unit{\kg\per\m\cubed}),\\$E$ (\unit{\Pa}), $\nu$} & $\hat{d}$ (\unit{\m}) & \makecell{$\mu$, $\epsilon_v$ (\unit{\m\per\s}), \\ friction iters.} &
        \makecell{Newton \\ tol. (\unit{\m})} \\
        \toprule
        \makecell{Armadillo-rollers\\(\cref{fig:high-order-teaser,fig:arma-roller-proxy})} & 0.025 & 1e3, 5e5, 0.2 & 1e-3 & 0.5, 1e-3, 1 & 1e-3 \\ \hline
        \makecell{Bending beam\\(\cref{fig:beam-bending})} & 0.1 & 1e3, 1e7, 0.4 & 1e-3 & 0.5, 1e-3, 10 & 1e-5 \\ \hline
        \makecell{Bouncing ball\\(\cref{fig:bouncing-ball})} & 0.001 & \makecell{700, 5.91e5,\\0.45} & 1e-3 & 0.2, 1e-3, 1 & 1e-12 \\ \hline
        \makecell{Mat-twist\\(\cref{fig:mat-twist})} & 0.04 & 1e3, 2e4, 0.4 & 1e-3 & - & 1e-5 \\ \hline
        \makecell{Balancing armadillo\\(\cref{fig:arma-balance})} & 0.1 & \makecell{$1e3^\star$, 1e11,\\0.2} & 1e-5 & 0.1, 1e-3, 20 & 1e-5 \\ \hline
        \makecell{Rolling ball\\(\cref{fig:rolling_ball_friction})} & 0.025 & 1e3, 1e9, 0.4 & 1e-3 & 1.0, 1e-3, $\infty$ & 1e-5 \\ \hline
        \makecell{Nut-and-bolt\\(\cref{fig:screw})} & 0.01 & \makecell{8050, 2e11,\\0.28} & 1e-4 & - & 1e-5 \\ \hline
        \makecell{Trash-compactor\\(\cref{fig:trash-compactor})} & 0.01 & 1e3, 1e4, 0.4 & 1e-3 & - & 1e-5 \\ \hline
        \makecell{Microstructure\\(\cref{fig:microstructure})} & 0.01 & \makecell{1030, 6e5,\\0.48} & 1e-5 & 0.3, 1e-3, 1 & 1e-4
    \end{tabular}
\end{table}

%% file: figs/sim_results.tex
\begin{table}
    \caption[High-order IPC summary of results]{Summary of results shown in \cref{sec:results}. For each example, we report the number of tetrahedra (\#T) used for elasticity, the number of surface triangles (\#F) used for collision processing, and the total running time of the simulation. Names correspond to the same given in each figure.}
    \label{tab:sim_results}
    \centering
    \footnotesize
    \begin{tabular}{cl|c|c|r}
        \multicolumn{2}{c|}{Scene} & \#T & \#F & Running Time \\
        \hline
        \multirow{4}{*}{\shortstack{Armadillo-rollers\\(\cref{fig:arma-roller-proxy})}}
        & Baseline & 386K & 24K & \dhms{2}{13}{19}{00} \\ %
        & \centering Ours$^\star$, $P_1$ & 1.8K & 24K & \ms{57}{36} \\ %
        & \centering Ours$^\dagger$, $P_2$ & 4.7K & 23K &\hms{3}{58}{00} \\ %
        & \centering Ours$^\ddagger$, $P_2$ & 4.7K & 24K &\hms{7}{14}{32} \\ %
        \hline
        \multirow{5}{*}{\shortstack{Bending beam\\(\cref{fig:beam-bending})}}
        & $P_1$ coarse        & 48   & 72   & 15s \\ %
        & $P_1$ reference     & 25K  & 4.4K & \ms{7}{43} \\ %
        & $P_2$               & 1.1K & 2.8K & 32s \\ %
        & $P_3$               & 48   & 5.5K & 58s \\ %
        & $P_1$ time budgeted & 48   & 5.5K & 57s \\ %
        \hline
        \multirow{4}{*}{\shortstack{Bouncing ball\\(\cref{fig:bouncing-ball})}}
        & Dense $P_1$ & 8.8K & 5.1K & \ms{4}{16} \\ %
        & Coarse $P_1$ & 30 & 32 & 8s \\ %
        & Dense Surface & 30 & 2.4K & 12s \\ %
        & $P_4$ & 30 & 2.4K & 26s \\ %
        \hline
        \multirow{3}{*}{\shortstack{Mat-twist\\(\cref{fig:mat-twist})}}
        & $P_1$ coarse & 2.2K & 1.6K & \ms{2}{47} \\ %
        & $P_1$ time budgeted & 54K & 37K & \hms{6}{7}{12} \\ %
        & $P_2$ & 2.2K & 129K & \hms{6}{19}{52} \\ %
        & $P_1$ & 230K & 141K & \dhms{2}{14}{13}{00} \\ %
        \hline
        \multirow{3}{*}{\shortstack{Balancing armadillo\\(\cref{fig:arma-balance})}}
        & Fine & 5.9K & 3.7K & 17s  \\
        & Coarse & 585 & 486 & 19s \\
        & Optimized & 585 & 486 & 9s\\
        \hline
        \multirow{2}{*}{\shortstack{Rolling ball\\(\cref{fig:rolling_ball_friction})}}
        & Baseline & 8.8K & 5.1K & \ms{5}{52}\\
        & Ours & 26 & 5.1K & 47s\\
        \hline
        \multirow{2}{*}{\shortstack{Nut and bolt\\(\cref{fig:screw})}}
        & Baseline & 6.1K & 5.2K & \ms{22}{04} \\
        & Ours & 492 & 5.2K  & \ms{9}{40} \\
        \hline
        \multirow{3}{*}{\shortstack{Trash-compactor\\(\cref{fig:trash-compactor})}}
        & Baseline & 21K & 8.6K &  \hms{5}{08}{25} \\
        & Ours & 3.5K & 8.5K  & \hms{2}{20}{16} \\
        & Ours, $P_2$ & 3.5K & 41K & \hms{24}{23}{00} \\
        \hline
        \multirow{2}{*}{\shortstack{Microstructure\\(\cref{fig:microstructure})}}
        & $P_1$ & 6.4K & 143K & \hms{6}{34}{09} \\
        & $P_2$ & 6.4K & 143K & \hms{6}{04}{48} \\
    \end{tabular}
\end{table}